\documentclass[useAMS,usenatbib]{mn2e}
\usepackage{graphicx}
\usepackage{ amssymb }
\usepackage{color}
\usepackage{rotating}
\usepackage{pdflscape}

\title[Discs in Chamaeleon I]{The Herschel view of circumstellar discs: a multi-wavelength study of Chamaeleon I}
\author[Rodgers-Lee et al.]{Donna Rodgers-Lee$^{1}$\thanks{E-mail:
donna@cp.dias.ie}, Alexander Scholz$^{1,2}$, Antonella Natta$^{1,3}$, Tom Ray$^{1}$\\ 
$^{1}$ School of Cosmic Physics, Dublin Institute for Advanced Studies, 31 Fitzwilliam Place, Dublin 2, Ireland \\
$^{2}$ School of Physics and Astronomy, University of St. Andrews, North Haugh, St Andrews, Fife KY16 9SS \\
$^{3}$ INAF - Osservatorio Astrofisico di Arcetri, Largo E. Fermi 5, I-50125 Firenze, Italy}
\begin{document}
\date{Accepted 2014 May 6. Received 2014 April 21; in original form 2014 January 15}

\pagerange{\pageref{firstpage}--\pageref{lastpage}} \pubyear{2013}

\maketitle

\label{firstpage}

\begin{abstract}

We present the results of a multi-wavelength study of circumstellar discs around 44 young stellar objects in the 3\,Myr old nearby Chamaeleon-I star-forming region. In particular, we explore the far-infrared/submm regime using Herschel fluxes. We show that Herschel fluxes at 160-500$\,\mu$m can be used to derive robust estimates of the disc mass. The median disc mass is 0.005$\,M_{\odot}$ for a sample of 28 Class IIs and 0.006$\,M_{\odot}$ for 6 transition disks (TDs). The fraction of objects in Chamaeleon-I with at least the `minimum mass solar nebula' is 2-7\%. This is consistent with previously published results for Taurus, IC348, $\rho$\,Oph. Diagrams of spectral slopes show the effect of specific evolutionary processes in circumstellar discs. Class II objects show a wide scatter that can be explained by dust settling. We identify a continuous trend from Class II to TDs. Including Herschel fluxes in this type of analysis highlights the diversity of TDs. We find that TDs are not significantly different to Class II discs in terms of far-infrared luminosity, disc mass or degree of dust settling. This indicates that inner dust clearing occurs independently from other evolutionary processes in the discs.

\end{abstract}


\section{Introduction}
\label{sec:intro}

Young stellar objects (YSOs) are an area of intense research as it is in the early stages of star formation that planetary systems are formed. The evolution of YSOs is often described using three main stages \citep{lada_1987,adams_1987}: Class I sources are thought to represent the earliest stage where the protostar is deeply embedded in an envelope of dust and gas which strongly accretes onto a circumstellar disc. Class II objects possess a circumstellar disc with lower accretion rates and the envelope has mostly disappeared. Class III objects no longer possess a circumstellar disc, or the disc is at least vastly dissipated. There have been further stages added such as Class 0 which is thought to be an earlier evolutionary phase than that of Class I \citep{andre_1993}. Another classification is that of flat spectrum sources which are thought to represent an evolutionary stage between Class I and Class II. 

The circumstellar discs around Class II objects evolve primarily due to viscosity, but also because of a number of other physical processes such as grain growth, dust settling, dissipation by photoevaporation and the formation of planets may take place. The evidence and effect of these physical processes are observable in the spectral energy distribution (SED) of YSOs \citep{watson_2007,dominik_2007,williams_2011}.

In this context, the frequently discussed transition discs  (TDs, see Section 7 in \citet{williams_2011} and references therein) play an important role. TDs were first identified from their SEDs which show a large deficit of emission in the near- and mid-infrared when compared to the median SED of Class IIs in the same star forming region. In some cases, millimetre interferometry data shows an inner cavity \citep[e.g.][]{brown_2007}. TDs are often considered to be a later evolutionary stage than Class II as the circumstellar disc is thought to have been cleared out by the formation of a planet \citep{rice_2003,quillen_2004}, but there are many other viable possibilities.  Stellar-mass companions \citep{jensen_1997,kraus_2008}, dust coagulation and evolution \citep{dullemond_2005,krauss_2007}, inside-out evacuation by the magnetorotational instability \citep{chiang_2007} or photoevaporation \citep{alexander_2006,armitage_2007} could also be responsible. The possibility of observing and understanding the early stages of planet formation makes TDs some of the most interesting YSOs to study. 

The availability of fluxes from the PACS and SPIRE instruments of the Herschel Space Observatory \citep{pilbratt_2010,poglitsch_2010,griffin_2010} gives us the opportunity to study for the first time the full SED for large numbers of YSOs. For Class II objects, the stellar photosphere is prominent from optical to near-infrared (NIR) wavelengths whereas the emission from circumstellar discs dominates in the mid-infrared (MIR) and far-infrared (FIR). The combination of 2MASS, DENIS, Spitzer, WISE, Herschel, LABOCA and mm data allows us to examine the entire emission spectrum of the discs. Depending on the wavelength range, different physical parameters can be investigated. In particular, the Herschel wavelength domain allows us to investigate the geometry of the outer discs and their disc masses. The inclusion of Herschel data also allows us to re-examine the classification of YSOs. 

In this paper we aim to investigate circumstellar discs using a multi-wavelength dataset, including Herschel fluxes, for objects in the star forming region Chamaeleon-I. The purpose of this study is to examine the discs using the full SED, to provide criteria to facilitate the study of larger samples, and to characterise possible targets for detailed follow-up observations, e.g., with the new submm/mm interferometer ALMA. We describe our sample of YSOs in Section~\ref{sec:description}. The construction and appearance of the SEDs is discussed in Section~\ref{sec:seds}. Section~\ref{sec:colour} illustrates spectral slope diagrams making use of the Herschel data. Section~\ref{sec:lum} investigates the fractional disc luminosities at Herschel wavelengths. We discuss the possibility of Herschel data tracing the mass of the disc and we present our results for the disc masses in Section~\ref{sec:disc_masses}. Section~\ref{sec:tds} focuses on fitting the SEDs of the TDs in our sample. Complementary information is given in the appendices.

\section{Description of sample}
\label{sec:description}

The Chamaeleon-I (Cha-I) star-forming region contains a well studied sample of YSOs, with a median age of $\sim$\,3\,Myr \citep{luhman_2007}, at a distance of $\sim$\,160\,pc \citep{whittet_1997}. It suffers little extinction which makes it an ideal region to study YSOs.
\citet{winston_2012} present Herschel data for 49 YSOs in Cha-I. The data are from the Herschel Gould Belt survey \citep{andre_2010} which has covered the entire Cha-I cloud. We will not consider in the following a B9.5 star (T32) and the object T54 which suffers from extended source contamination \citep{matra_2012}. We exclude ESO H$\alpha$ 569 which we confirm to be an edge-on disc (see Appendix~\ref{subsec:edge-on}). In addition, we exclude ISO 237 and Hn10e which suffer from contamination (see Appendix~\ref{subsec:edge-on}). For a full list of the 44 objects in our sample and their derived properties see Table~\ref{table:entire_sample}.

The temperatures, luminosities and extinctions ($A_J$) were obtained for 35 out of our 44 objects from \citet{luhman_2007}. The stellar properties are used in Section~\ref{sec:lum}, i.e. only 35 objects are considered in this section. The size of the sample being considered in each section will be listed at the beginning of the section for clarity.

\begin{figure}
\center
 \includegraphics[width=0.50\textwidth]{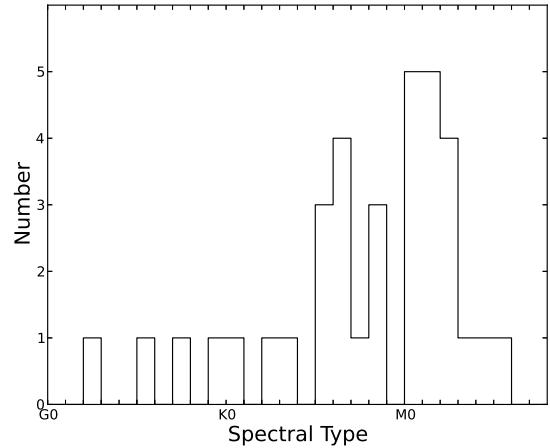}
 \caption{Distribution of different spectral types \citep{luhman_2007} for our sample.
 \label{fig:hist_spec}}
\end{figure}

The temperature range of our sample is from $\sim 3090 - 5860$\,K and the spectral types range between $\mathrm{G2} - \mathrm{M5.25}$, as shown in Fig.~\ref{fig:hist_spec}. This means the sample covers a mass range of $\sim 0.1 - 2.0 \ \mathrm{M}_\odot$. From the list in \citet{winston_2012} we also take the SED classification for these objects. They comprise 29 Class IIs, 6 flat spectrum (FS) sources, 2 Class II/TD, 1 TD, 4 Class Is, 1 Class 0 and 1 Class III. The original source of these classifications is \citet{luhman_2008}. Note that these classes are based on Spitzer data only (i.e. up to $\lambda = 24\,\mu$m).

Recently \citet{ribas_2013} published Herschel data for 9 objects, 8 of which are in our sample. Their fluxes are somewhat different (on average 18\% higher at 70$\,\mu$m) than in \citet{winston_2012}, possibly due to different data reduction techniques. We also note that 12 of the sources analysed here are also contained in the 17-object sample discussed in a recent paper by \citet{oloffson_2013}. They focus on the low mass stars and brown dwarfs in Cha-I, consequently all 12 are M dwarfs and at the low-mass end of the spectrum covered in our paper. \citet{oloffson_2013} analyse the disc SEDs with Herschel/PACS data using radiative transfer modeling, and some of their results are complementary to ours.

\begin{figure}
\centering
 \includegraphics[width=0.53\textwidth]{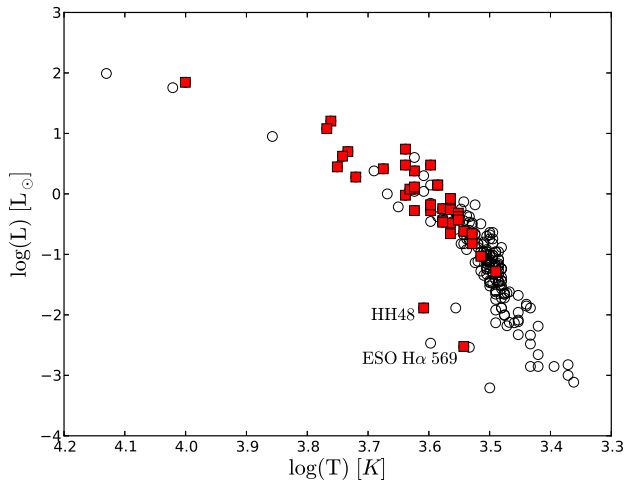}
 \caption{HR diagram of Cha-I: Open circles represent all known YSOs in Cha-I, from \citet{luhman_2007}, red squares are the objects in our sample. The two underluminous outliers are labelled. ESO H$\alpha$\,569 is one of the objects excluded from our sample. }
\label{fig:hr}
\end{figure}

Our list of objects is not a complete sample of all YSOs in Cha-I. First, we are only analysing objects detected by Herschel. From \citet{winston_2012} the Herschel detection limit was adopted to be $\sim $0.1\,Jy at all bands which corresponds to a spectral energy density detection limit of $4.3 \times 10^{-12}\mathrm{erg \ sec}^{-1} \ \mathrm{cm}^{-2}$ at 70\,$\mu$m. Objects with discs fainter than this detection limit are missing if they exist. Second, as discussed above, 9 objects do not have stellar properties listed in \citet{luhman_2007} and are excluded from the analysis in Section~\ref{sec:lum}. The luminosity bias in the far-infrared also results in missing most of the very low mass members of Cha-I. This is obvious from the Hertzsprung-Russell (HR) diagram of all the known objects in Cha-I \citep{luhman_2007}, as shown in Fig.~\ref{fig:hr}. In this diagram, the objects in our sample are marked in red. The two outliers from our sample in the HR diagram are ESO H$\alpha \ 569$ and HH48. These 2 objects are discussed further in Appendix~\ref{subsec:edge-on}, along with their SEDs. 

\section{Spectral Energy Distributions}
\label{sec:seds}

\indent We constructed SEDs for the sample using DENIS, 2MASS, WISE, Spitzer, Herschel, LABOCA and 1300\,$\mu$m data where available, see Table~\ref{table:photometry_list} for a complete list of the photometry used. 23 out of the 44 objects have 870\,$\mu$m or 1.3\,mm data. We also used Spitzer IRS \citep{irs_2004} data where available. The extinction law from \citet{cardelli_1989} with $R_V = 3.1$ was used to deredden the data points. 

Some of the objects in our sample were observed by \citet{belloche_2011} using LABOCA, which has an angular resolution of 19.2$\arcmin \arcmin$. We took any matches with 2MASS coordinates within $\sim 5 \arcmin\arcmin$. Seventeen objects were matched, but with this criterion, 2 objects are clearly misassociated (ESO H$\alpha$\,559 and 2MASS J11064658-7722325). This was decided by looking at the full SEDs for these objects and the LABOCA flux density for these objects was not used. 

\subsection{Modeling the photospheric SED}
\label{subsec:stellar_atmos}

The first aim was to correctly determine the stellar photosphere component from the entire SED. This made it possible for us to examine the dust component of the SED to compute disc luminosities and compare it with models. The stellar component of the SED was constructed using BT-NextGen\footnote{http://phoenix.ens-lyon.fr/simulator/index.faces} model atmospheres from \citet{allard_2012} with $\mathrm{log} \ \mathrm{g} =  3.5,$ for all objects. The flux densities given by the model atmospheres were multiplied by a dilution factor of $(R_{star}/D)^2$, where $R_{star}$ is the radius of the star and $D$ is the distance to the star (160 pc). The radius of the star was calculated using the temperature and luminosity given in \citet{luhman_2007}. There were no further adjustments made to fit the SEDs (e.g., no normalisation). The NIR data were used to check that the model atmosphere chosen for the object fit the data. For 30 out of the 35 objects, the photospheric model fit the datapoints reasonably well, which confirms the validity of the assumed stellar properties. The remaining 7 show significant deviations, most likely due to variability. The optical and NIR photometry in the literature for Cha-I originates from observations spanning more than two decades. On these timescales, most YSOs show variations up to 30\%, and a subsample even more \citep{scholz_2012}. For the 7 objects in question we obtained a new set of photometry, which is discussed in Appendix~\ref{sec:var_sources}. All seven are confirmed as variable stars, two of them with large amplitudes of $>1$ mag.

\begin{table*}
\centering
\caption{List of photometry sources and relevant wavelengths }
\begin{tabular}{@{}lcllll@{}}
\hline
Telescope/Survey & Instrument & $\lambda [\mu\mathrm{m}]$/band & Ref. & Zero-point flux [Jy] & Ref. \\
\hline
-	 & -			& R, V		& 1		& 2870.0, 3540.0 & 11 \\
HST 	 & - 			& 0.56  	& 2		&  - & - \\
DENIS 	 & -	 		& I, J, K 	& 3		& 2499.0, 1595.0, 665.0 & 3 \\
2MASS 	 & -	 		& J, H, K 	& 4		& 1594.0, 1024.0, 666.7 & 12 \\
WISE	 & -	        	& 3.4, 4.6, 12.0, 22.0 &5 	& 309.540, 171.787, 31.674, 8.363 & 13\\
Spitzer  & {\it IRAC} 	& 3.6, 4.5, 5.8, 8.0  &	6	& 280.9, 179.7, 115.0, 64.13 & 14\\
         & {\it MIPS}		& 24 		& 6		& 7.17 &  14\\
Herschel & {\it PACS} 	& 70, 160 	& 7		& - & - \\
	 & {\it SPIRE} 		& 250, 350, 500 & 7 		& - & -  \\
APEX   & LABOCA	 		& 870  		& 8,9		& - & -  \\
-  & -				& 1300  	& 10		& - & -  \\

\hline 
\end{tabular}

$^1$ \citet{strom_1992}, $^2$ \citet{robberto_2012}, $^3$ \citet{fouqu_2000}, $^4$ \citet{skrutskie_2006}, $^5$\citet{wise_2010} ,\\
 $^6$ \citet{luhman_2008}, $^7$ \citet{winston_2012}, $^8$ \citet{belloche_2011}, $^9$ \citet{cieza_2013}, $^{10}$ \citet{henning_1993},\\ $^{11}$ \citet{skinner_1996}, $^{12}$ \citet{2MASS},$^{13}$\citet{WISE},$^{14}$ http://irsa.ipac.caltech.edu/Missions/spitzer.html \\
\label{table:photometry_list}
\end{table*}

\subsection{SEDs for typical YSOs}
\label{subsec:discuss_seds}

\begin{figure*}
\centerline{\includegraphics[width=0.35\textwidth]{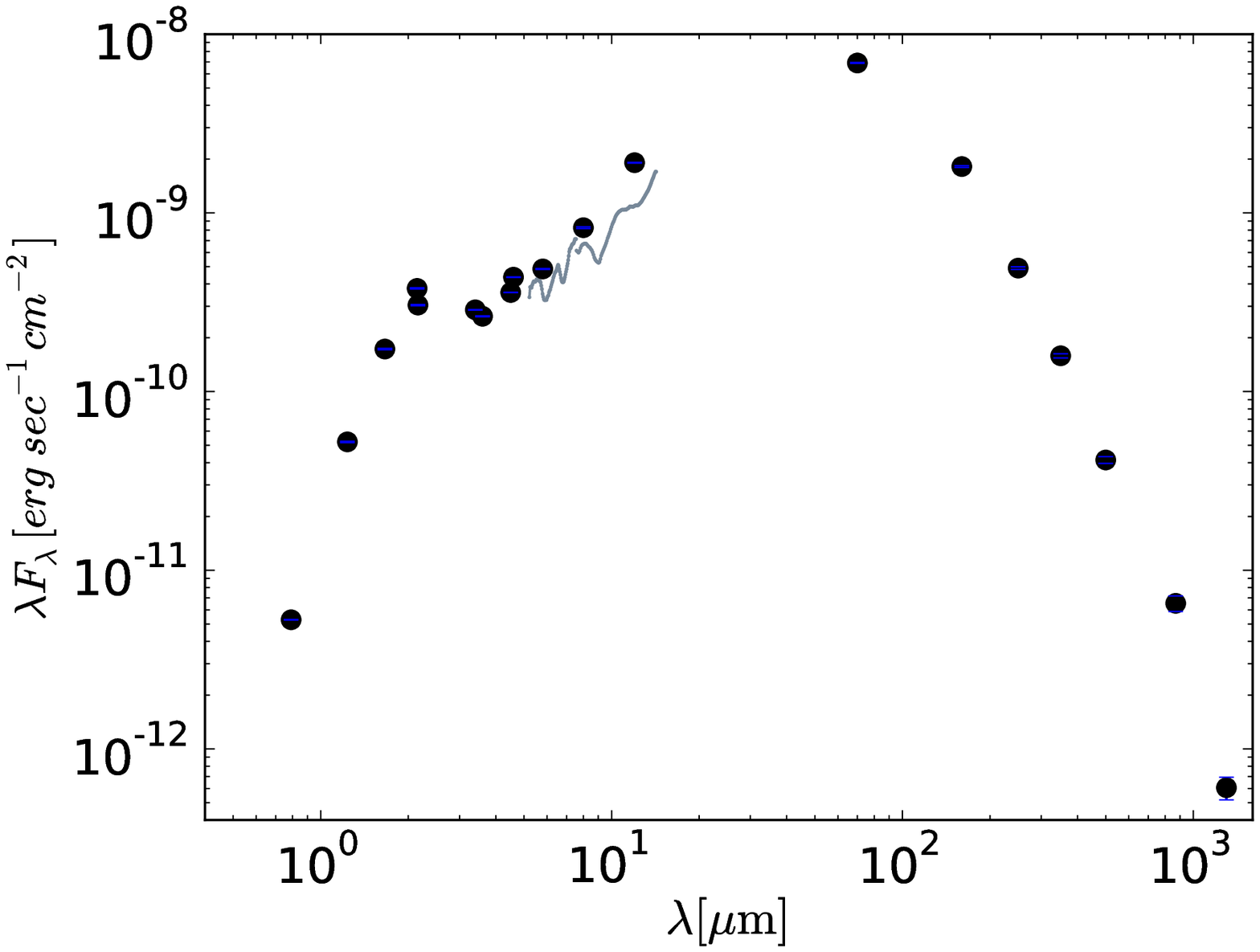}\qquad
\includegraphics[width=0.35\textwidth]{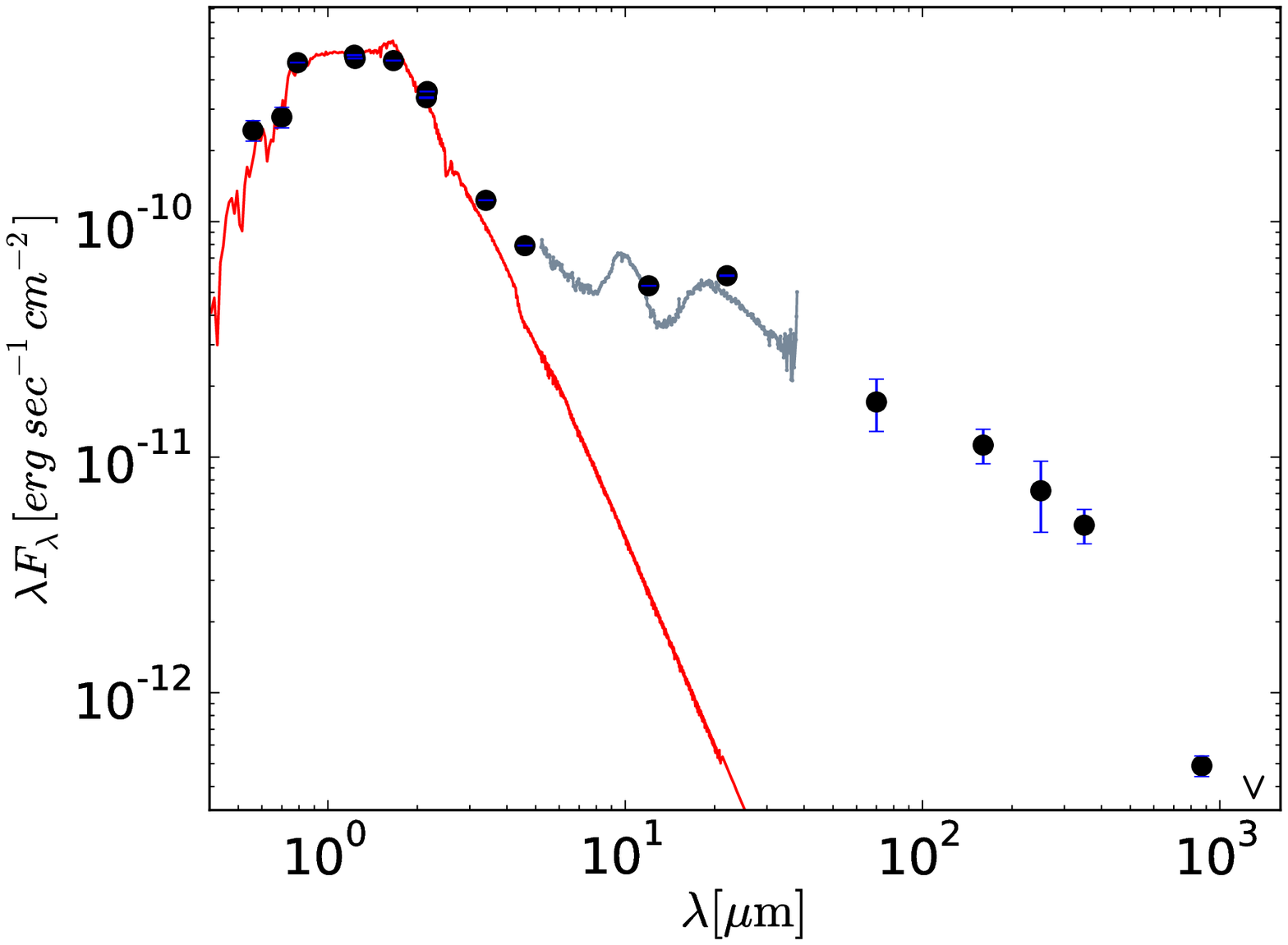}\qquad
\includegraphics[width=0.35\textwidth]{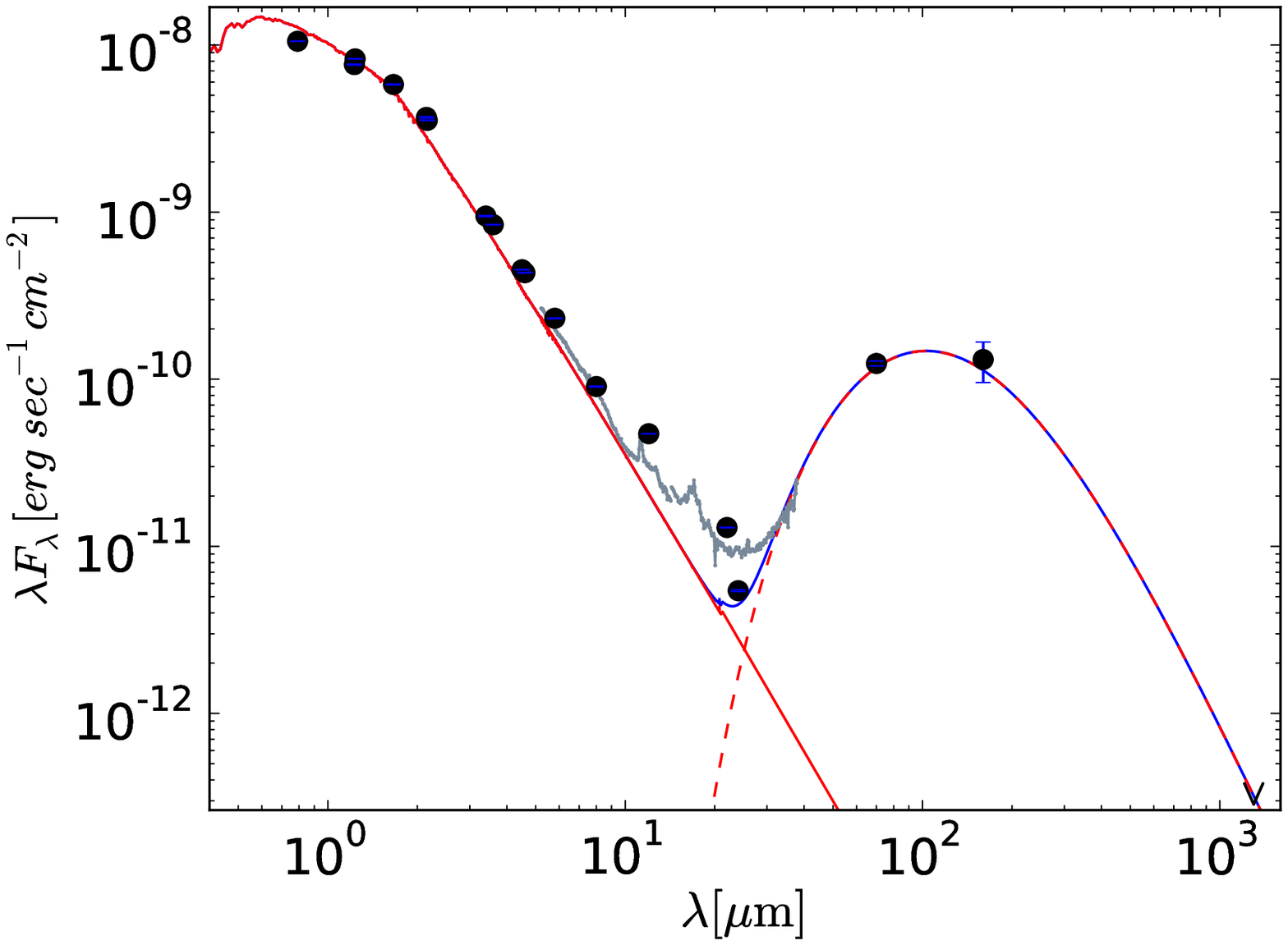}}
\centerline{(a) IRN/P24, Class I\hspace{0.25\textwidth} (b) SY Cha, Class II \hspace{0.25\textwidth} (c) T21, TD}

\caption{Variety of SEDs for our sample. The red lines are the model atmospheres from \citet{allard_2012}. For T21 the dotted red line is the simple model fit (see Appendix~\ref{subsec:model}) and the blue line is combination of both.
\label{fig:seds}}
\end{figure*}

The SEDs in our sample show great diversity. In Fig.~\ref{fig:seds} we show 3 examples from our sample illustrating a typical Class I, Class II and TD. Objects similar to Fig.~\ref{fig:seds}\,(a) with a rising SED through the NIR and MIR are identified in the literature as Class 0 or Class I. This strong excess cannot be explained by a disc alone and is at least partly caused by a circumstellar envelope. The majority of the sample resembles Fig.~\ref{fig:seds}\,(b) with excess at all IR wavelengths, but declining SEDs. These are typical Class IIs, the IR emission is consistent with being produced by a circumstellar disc.

A third type of SED is shown in Fig.~\ref{fig:seds}\,(c). This object, T21, shows negligible excess emission above the photospheric level in the NIR to MIR. It has steeply increasing flux densities around 10-20\,$\mu$m, and beyond 70\,$\mu$m an SED similar to Class IIs. This is the typical appearance of TDs. It is usually explained by an opacity hole in the inner disc (see Section~\ref{sec:intro}). This object is categorized as Class III by \citet{luhman_2008} based on the photometry in the mid-infrared, but was already noted as TD by \citet{kim_2009} using the IRS spectrum (5-38$\,\mu$m). 

These three objects represent three different stages in disc evolution. However, there are other objects which are not as easily classified. In Section~\ref{sec:colour}, using Herschel data, we demonstrate that there is a continuum of objects between these clearly distinct examples of Class I, Class II and TD.

\section{Spectral slopes and disc evolution}
\label{sec:colour}

Spectral slopes were examined, making use of the Herschel data, to classify and characterise the objects in our sample. In the following discussion the spectral slope is defined as \citep{adams_1987}:

\begin{equation}
\hspace{10mm} \alpha_{\lambda_1 -  \lambda_2} = \frac{\rmn{log}(\lambda_1 F_{\lambda_1}) - \rmn{log}(\lambda_2 F_{\lambda_2})}{\rmn{log}(\lambda_1) - \rmn{log}(\lambda_2)}
\label{eq:spec_slope}
\end{equation}

\begin{figure}
\centering
 \includegraphics[width=0.53\textwidth]{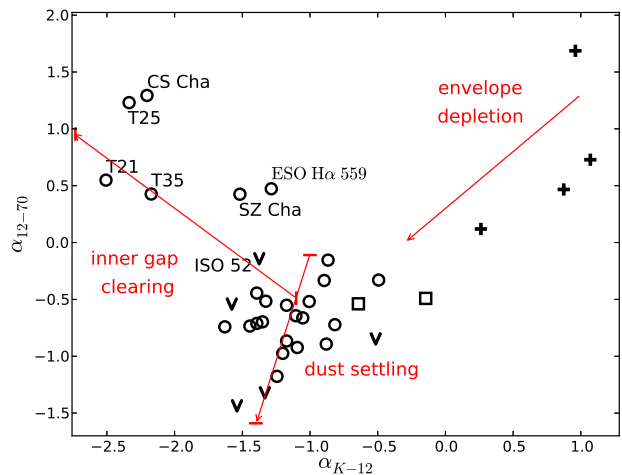}
  \caption{Spectral slope diagram of entire sample: `$+$' are Class Is, `$\square$' are flat spectrum sources, `$\circ$' are Class IIs and TDs,`$\vee$' are upper limits. Objects that will be discussed in Section~\ref{sec:tds} are labelled. }
\label{fig:k-12-12-70}
\end{figure}

\begin{figure}
\centering
 \includegraphics[width=0.53\textwidth]{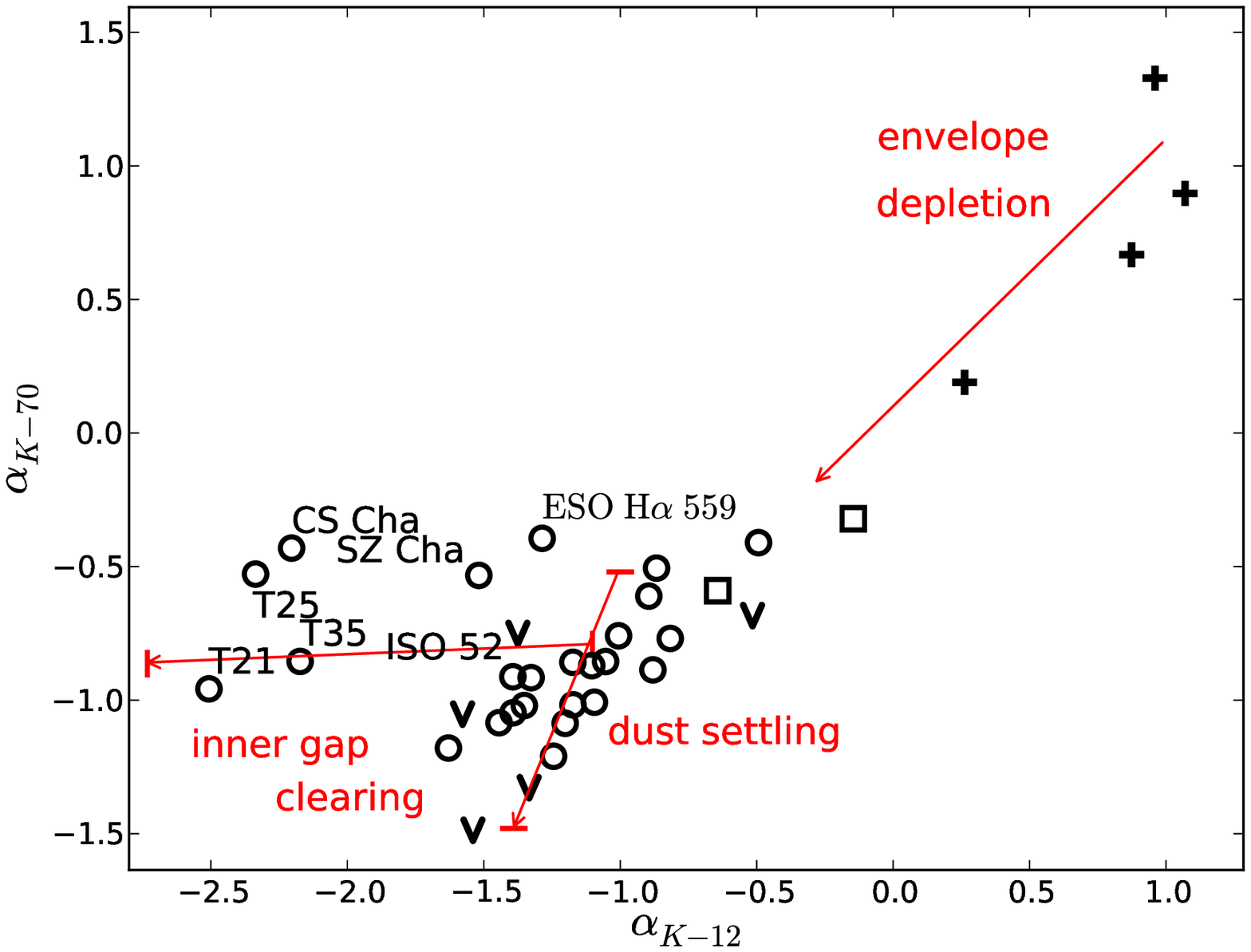}
 \caption{Spectral slope diagram of entire sample: with symbols as in Fig.~\ref{fig:k-12-12-70}} 
\label{fig:col-k-12-70}
\end{figure}

We have plotted $\alpha_{12-70}$ against $\alpha_{K-12}$ in Fig.~\ref{fig:k-12-12-70} and $\alpha_{K-12}$ against $\alpha_{K-70}$ in Fig.~\ref{fig:col-k-12-70} as possible ways to (a) separate the various different classes of objects and (b) to illustrate the effects of the physical processes that affect the discs. The classifications used in the figures come from \citet{luhman_2008} for the Class Is and flat spectrum sources. We have not labelled the known TDs and Class IIs differently. Both diagrams show a clear `Class II locus' around (-1,-1), where the bulk of our sample is located. 

In both diagrams three broad trends are visible. First, there seems to be a continuous trend from the upper right (where the Class I objects appear) to the Class II locus. This slope reflects the drop in the high levels of NIR/MIR emission observed in the SED and the evolution from rising SEDs in the Class Is to declining SEDs in the Class IIs. This change cannot be explained with an evolution of the discs themselves. The most likely explanation is the gradual disappearance of the circumstellar envelope. The arrow labelled `envelope depletion' is merely indicative of the effect we believe that envelope depletion would have and does not correspond to any model data points. The approximate line between Class I and flat spectrum/Class IIs is at $\alpha_{K-12}, \alpha_{K-70} \approx 0$, but the diagrams do not show a clear separation between Class I and Class II. 

Second, in both spectral slope plots, the Class IIs show a broad scatter of about 1 order of magnitude on both axes. The scatter is most likely caused by the evolution from flared to flat discs, possibly caused by dust settling \citep{dullemond_2004,dalessio_2006}. We investigated the changes in the SEDs due to settling using 2-layer disc models \citep{chiang_1997,chiang_1999,dullemond_2001} with different dust populations in the surface and midplane of the disks, as described in \citet{ricci_2010}, where the details of the adopted dust model can also be found. The small grains on the disk surface have a dust-to-gas ratio of 0.01 in fully flared models. We describe the settling by decreasing this ratio by a constant factor $\epsilon$, as in \citet{dalessio_2006}, from 1 (fully flared models which is the upper dash on the arrows in Figs.~\ref{fig:k-12-12-70},~\ref{fig:col-k-12-70}) to 0.1, 0.01, 0.001 (the lower dash on the arrows in Figs.~\ref{fig:k-12-12-70},~\ref{fig:col-k-12-70}) respectively. Using this model we found that a fully flared disc around a Class II source with characteristics typical for our sample (SY Cha in Fig.~\ref{fig:seds}) sits at the upper right corner of the Class II `clump' in both diagrams. On the other hand, a Class II with an almost flat disc sits at the lower left corner of the Class II `clump'. This indicates that all the scatter in $\alpha_{12-70}$ and $\alpha_{K-70}$ and a large part of the scatter in $\alpha_{K-12}$ can be explained by varying the degree of flaring. Conversely, if settling is indeed the physical underlying process the scatter can be explained by varying its efficiency. An arrow in both diagrams, which corresponds to the position and spans the range of model data points, indicates the direction of increasing dust settling. \citet{oloffson_2013} find a distribution in the amount of flaring in their sample of M dwarfs from Cha-I, similar to our result.

Seven objects in these diagrams are labelled (SZ Cha, CS Cha, T21, T25, T35, ESO H$\alpha$ 559 and ISO 52). These labelled objects are located above (in Fig.~\ref{fig:k-12-12-70}) and to the left (in Fig.~\ref{fig:col-k-12-70}) of the Class II locus. For this group $\alpha_{12-70}$ is higher than for the Class II group, but $\alpha_{K-12}$ is lower, i.e. there is evidence for a flux deficit in the NIR/MIR. Therefore, all these objects should be considered to be TDs. While the most extreme outliers in this group are well-known TDs (e.g. T21, see Section~\ref{subsec:discuss_seds}), others occupy the colour space between these prototypical TDs and Class II. For a more detailed discussion and the SEDs of these labelled objects, see Section~\ref{sec:tds} and Fig.~\ref{fig:seds_td}.

\citet{ribas_2013} use an analogue of Fig.~\ref{fig:k-12-12-70} to introduce a sharp distinction between TDs and Class II discs. In our figure, this separation is seen at:
\begin{eqnarray}
 &\alpha_{12-70} &\gtrsim 0  \mbox{ for TD objects} \nonumber \\
 &\alpha_{12-70} &\lesssim 0  \mbox{ for Class II objects}
\end{eqnarray} 
However, we note that there is no obvious gap in the diagram between Class IIs and TDs, thus, the spectral slope threshold given above is arbitrary and can only be used to classify the most extreme objects. Both diagrams show a continuum of datapoints from the Class II locus to the typical TDs. Objects like SZ Cha and ISO 52 (for their SEDs see Fig.~\ref{fig:seds_td}) are located in this transition area. This smooth transition from Class II to TD is also visible in Fig.~\ref{fig:col-k-12-70}.

Using the simple disc model of \citet{beckwith_1990} which is described in Appendix~\ref{subsec:model}, we aim to explain the difference between Class IIs and TDs in the spectral slope diagrams by assuming that the TDs are affected by inner disc clearing. We adopt plausible parameters for the central source and the global properties of the discs. These plausible parameters are the fiducial values mentioned in Appendix~\ref{subsec:model}. We only vary the inner radius of the discs, from about 0.4\,AU to 1.5\,AU. In Figs.~\ref{fig:k-12-12-70} and ~\ref{fig:col-k-12-70} an arrow illustrates the effect of an increasing inner disc radius on the spectral slope values. The arrow spans the range of the inner radii used in the model. Clearly, the presence of inner holes of varying sizes alone can broadly explain the phenomenological difference between Class IIs and TDs. For a more detailed discussion of the inner hole sizes and the nature of these gaps, we refer the reader to Section~\ref{sec:tds}.

The spread in the TDs $\alpha_{K-70}$ spectral slope is similar to that of the Class IIs. This slope is mostly unaffected by the inner disc clearing. The spread can again be explained by varying degree of dust settling, as for the Class IIs (see above). Thus, dust settling affects Class IIs and TDs in a similar way. Based on this finding, it seems plausible that the clearing of the inner disc, which may be due to a variety of processes, occurs independently from the settling process in the discs. This is in line with one of the conclusions in the study by \citet{sicilia-aguilar_2013b}, based on NIR and MIR data.

\begin{figure}
\centering
 \includegraphics[width=0.53\textwidth]{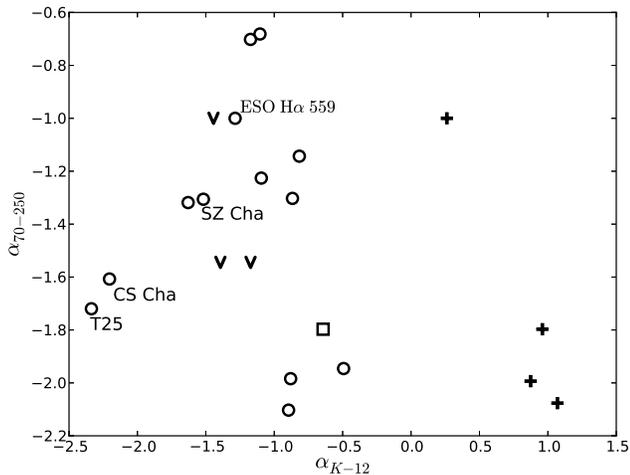}
 \caption{Spectral slope diagram of entire sample: with symbols as in Fig.~\ref{fig:k-12-12-70}} 
\label{fig:k-12-70-250}
\end{figure}

We also examined spectral slope diagrams using the data at $\lambda \ge 70\,\mu$m, but found no noticeable difference between TDs and the rest of the sample. As an example, we plotted $\alpha_{70-250}$ against $\alpha_{K-12}$ in Fig. \ref{fig:k-12-70-250}. At these long wavelengths, the spectral slopes of Class Is, Class IIs, and TDs are well mixed. Thus, the bulk of the dust in the outer disc, which determines the flux densities and spectral slopes at the submm/mm wavelengths, remains unaffected by whatever process causes the clearing in the inner part of the disc.

Note that Class IIIs and sources with only photospheric emission or very little infrared excess (e.g. from debris discs) would appear in the lower left corner of the diagrams which are not populated by our sample due to the lack of sensitivity in the Herschel survey.

\section{Disc Luminosities}
\label{sec:lum}

The luminosity of each disc was calculated to examine possible trends of luminosity with object class. Nine objects were omitted (which includes all 4 Class I objects) for the reasons discussed in Section~\ref{sec:description}, i.e. a sample of 35 objects remains.

The stellar luminosities were computed using the flux densities from the model atmospheres so that the stellar luminosity can be subtracted from the total luminosity of each object. $L_{star}(\lambda_1, \lambda_2)$ is the stellar luminosity, between $\lambda_1=0.1\,\mu$m and $\lambda_2=1000\,\mu$m, which was calculated by integrating over the model atmosphere flux density values and multiplying by $4 \pi D^2$, where $D$ is the distance to the star. The interpolation between data points is a straight line. Our luminosities were compared with the values from \citet{luhman_2007}; 95\% were within 10\% of these luminosities and 64\% were within 5\%. Given the different method used to compute the stellar luminosities our values are acceptable. 

The disc luminosity, $L_{disc}(\lambda_1,\lambda_2)$, between two particular wavelengths, $\lambda_1$, $\lambda_2$, was then calculated as follows:

\begin{equation}
\hspace{10mm} L_{disc}(\lambda_1, \lambda_2) = L_{both}(\lambda_1,\lambda_2) - L_{star}(\lambda_1,\lambda_2)
\end{equation}

\noindent where $L_{both}(\lambda_1,\lambda_2)$ is the total luminosity (the stellar and disc contributions combined) between $\lambda_1$, $\lambda_2$, calculated by integrating between the flux density data points for these wavelength values. In the FIR regime the stellar luminosity is small in comparison to the luminosity of the disc so $L_{disc}(\lambda_1, \lambda_2)$ will be most accurate there. In the NIR $L_{disc}(\lambda_1, \lambda_2)$ is very sensitive to the exact value of the photospheric contribution; it is difficult to quantify any disc contribution to the overall luminosity at these shorter wavelengths. We therefore do not examine disc luminosities for $\lambda < 8\,\mu$m. The fractional disc luminosity is defined as $L_{disc}(\lambda_1,\lambda_2)/L_{star}$, where $L_{star}$ is the total stellar luminosity. 

\begin{figure*}
\centerline{\includegraphics[width=0.53\textwidth]{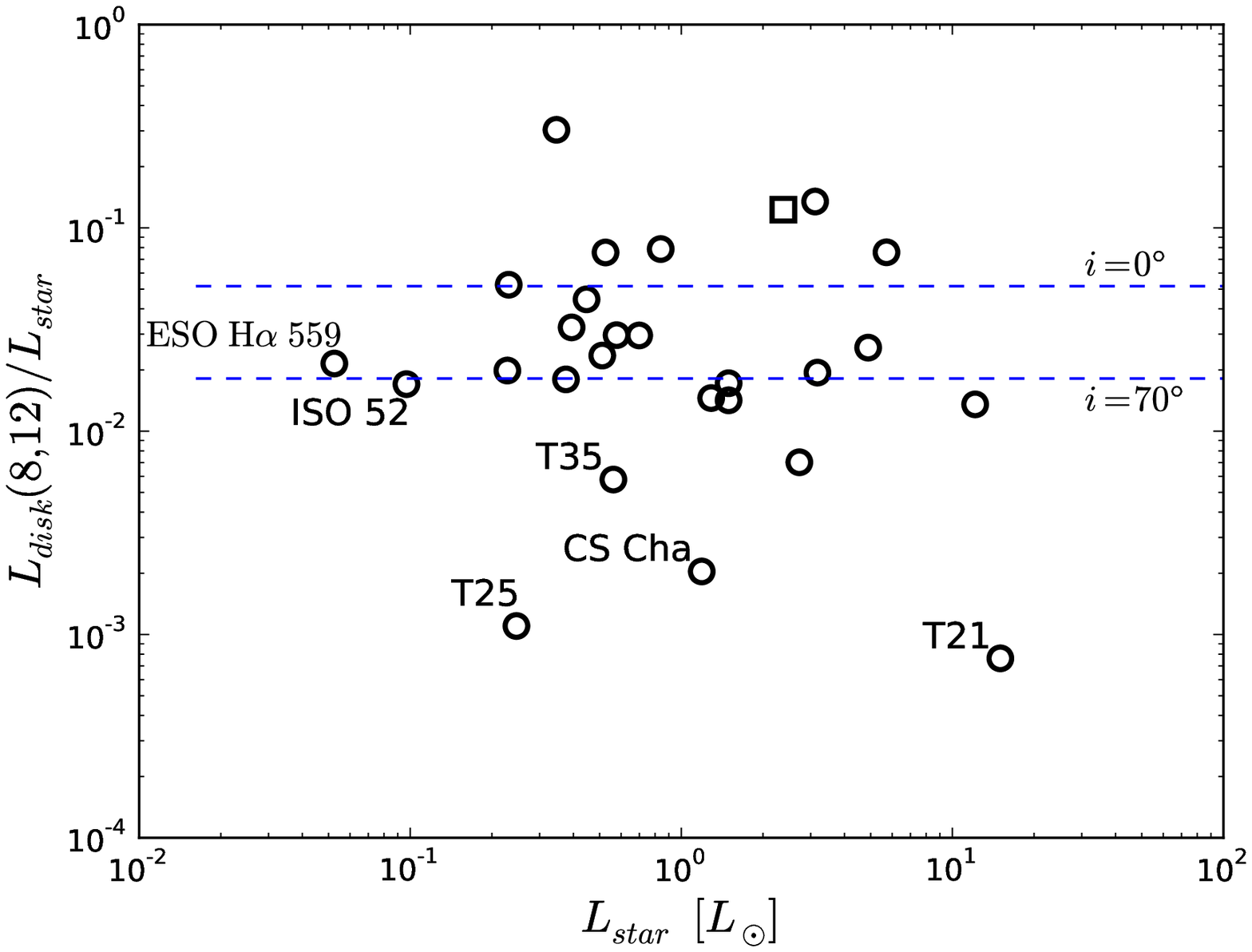}\qquad
\includegraphics[width=0.53\textwidth]{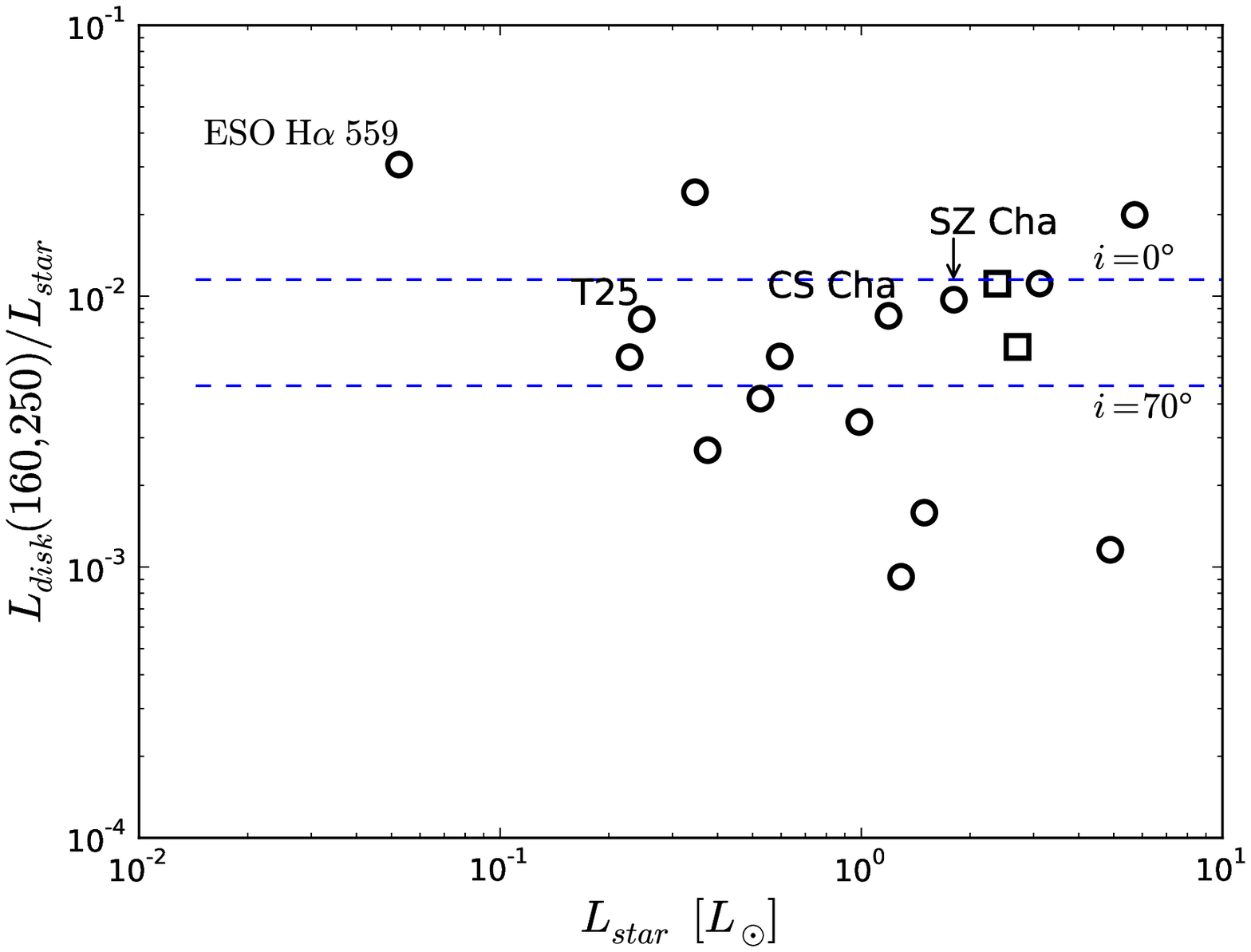}}
\centerline{(a)\hspace{0.53\textwidth} (b) }
\caption{Luminosity plots with symbols as in Fig.~\ref{fig:k-12-12-70}. The blue dotted lines are the flux density values for a typical Class II obtained from the simple model (see Appendix~\ref{subsec:model}) for different angles of inclination: $i = 0^\circ,70^\circ$.
\label{fig:i-lum}}
\end{figure*}

We plotted various combinations such as $L_{disc}(8, 12)/L_{star}$ and $L_{disc}(160,250)/L_{star}$, as shown in Figs.~\ref{fig:i-lum}\,(a),\,(b). In these plots, the same objects as in Section~\ref{sec:colour} are labelled. As pointed out in Section~\ref{sec:colour}, all these objects can be considered to be TDs with evidence of inner disc clearing.

These figures show no correlation between fractional disc luminosity and total stellar luminosity. Fig.~\ref{fig:i-lum}\,(a) indicates that most TDs are underluminous between $8-12\,\mu$m which is to be expected. However, in Fig.~\ref{fig:i-lum}\,(b) most of them are indistinguishable from other Class IIs. This implies that in the FIR the energy output of TDs is the same as Class IIs, confirming that inner disc clearing is independent from the evolution of the outer disc.

One of the objects that we have classed as a TDs, ESO H$\alpha$\,559, is overluminous in comparison to the rest of the TDs at the Herschel wavelength regime (see Fig.~\ref{fig:i-lum}, both panels). There is no evidence that this object is seen edge-on, i.e. inclination effects are not a plausible explanation for this anomaly. One option is that this a strongly flared disc with only a small inner hole.

In contrast to what Fig.~\ref{fig:i-lum}\,(b) shows (that the TDs are well mixed) \citet{ribas_2013} found that the TD flux densities at 70 and 160$\,\mu$m were larger than the median flux density of Class IIs. A combination of the following three effects could potentially resolve this discrepancy: (1) \citet{ribas_2013} are examining a sample of objects from Cha-I and Cha-II whereas our sample is only sources from Cha-I. (2) As mentioned in Section~\ref{sec:description} the flux densities reported in \citet{ribas_2013} are systematically higher than the flux densities in \citet{winston_2012} for the TDs, which we are using. \citet{ribas_2013} do not list the flux densities for the Class IIs, i.e. we cannot decide on this particular issue. (3) We are also examining the luminosity between 160-250$\mu$m and not between 70-160$\mu$m. We also note that the samples are quite small, i.e. it is difficult to establish
significant differences.

The disc luminosity is $\sim 0.1-0.2\,L_{star}$ for our sample between 8 and 500\,$\mu$m. In comparison, debris discs, which are the discs formed from remnant discs by planetesimal collisions, have been found to be $\sim 10^{-4} L_{star}$ \citep{rebull_2008}, which makes the most luminous circumstellar discs in Cha-I $\sim 10^3$ times more luminous than these debris discs and therefore clearly primordial. This is also not taking into account any contribution to the disc luminosity in the NIR regime.

The observed luminosities were compared with values obtained from our simple model (see Appendix~\ref{subsec:model}), for different inclination angles, $i = 0^\circ, 70^\circ$. SY Cha as a template for the stellar parameters, as it appears to be a typical Class II source, see Fig~\ref{fig:seds}\,(b). These model luminosities are overplotted in Figs.~\ref{fig:i-lum}\,(a)\,-\,(b). 
It is evident that the scatter in our luminosity plots is larger than expected from variation of the inclination angle. The scatter expected from 78\% of our objects should be half an order of magnitude but we have at least an order of magnitude scatter in our plots. This indicates that additional factors, such as dust settling, must be contributing to the diversity as already pointed out in Section~\ref{sec:colour}.

\section{Disc Masses}
\label{sec:disc_masses}

One of the main benefits of the submm/mm regime for studies of discs is that the dust becomes optically thin, so it is possible to infer the total amount of dust in the disc. In the following we will explore the possibility of calculating disc masses for the sample of discs in Cha-I using Herschel data.

\subsection{Testing `monochromatic' disc masses}
\label{subsec:disc_mass1}

Disc masses of a large number of objects are often computed from single wavelength sub-mm and/or mm fluxes. It is assumed that the emission is optically thin and that a single temperature (typically $\sim 20$\,K) can be assumed to describe the emission of the whole disc \citep{hildebrand_1983,beckwith_1990}:
\begin{equation}
\hspace{30mm} M_d = \frac{D^2 F_\nu}{\kappa_\nu B_\nu(T_c)}
\label{eq:m_d}
\end{equation}
Here $D$ is the distance to the star, $\kappa_\nu$ is the opacity, $F_\nu$ is the flux density and $B_\nu(T_c)$ is the Planck function at a characteristic temperature $T_c$. 

Disc masses calculated from this simple scaling law are called `monochromatic' masses in the following. So far, most surveys of disc masses have been carried on at 850$\,\mu$m or 1.3\,mm. The availability of the Herschel surveys of star-forming regions at shorter wavelengths (70-500$\,\mu$m) raises the question of how suitable these wavelengths are for estimating disc masses.

This question is addressed by using 8 objects as test cases in the sample. For these objects both Herschel data and  measurements of fluxes at longer wavelengths, either 850 or 1300$\,\mu$m, were available. The FIR/submm/mm SED of these objects is modelled with the simple model described in Appendix~\ref{subsec:model} for the stellar parameters listed in Table~\ref{table:entire_sample}. For each object a value of the disc mass,$M_d$, is derived. A fit of the SED is not carried out. Instead, the parameters of the model were varied within the plausible ranges until a satisfactory by-eye match was achieved. Note that we did not attempt to fit the NIR and MIR fluxes as this part of the SED does not affect the disc mass. In all 8 cases, a well-matched model is found with disc masses between 0.002 and 0.05$\,M_{\odot}$, called the `true' disc masses in the following. 

These `true' masses ($M_d$) are compared to the monochromatic disc masses ($M_d(\lambda)$) obtained from single-wavelength fluxes following Eq.~\ref{eq:m_d}. For the monochromatic masses, we adopt the same opacity law as for the true masses with $\beta = 1.0$
and $\kappa_{f=\mathrm{230GHz}} = 0.023\,\mathrm{cm}^2 \ \mathrm{g}^{-1}$ (see Appendix~\ref{subsec:model}). We also adopt a temperature of $T_c = 20 $\,K, a distance of $D = 160$\,pc, and a dust-to-gas ratio of 1:100. These parameters are chosen to be both plausible and easily comparable with a wide range of literature surveys (see for example \citet{beckwith_1990,shirley_2000,andrews_2005,scholz_2006,ricci_2010,lee_2011}). 

\begin{figure}
\centering
\includegraphics[width=0.53\textwidth]{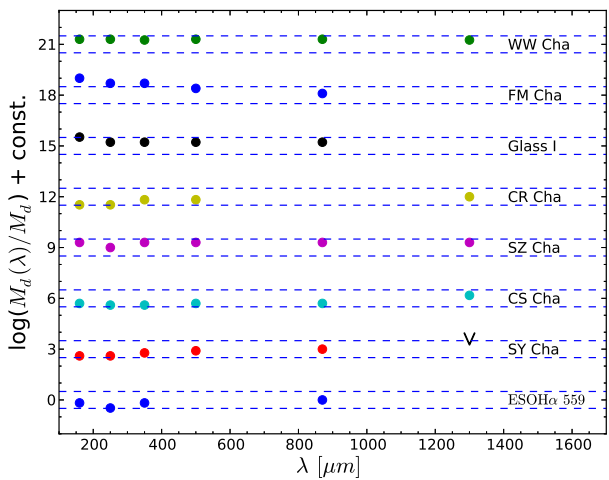}
 \caption{Ratio of monochromatic disc masses, $M_d(\lambda)$,  to `true' disc masses, $M_d$, from SED fitting for 8 objects in Cha-I, as a function of wavelength. Each object is shifted vertically and labelled; the dashed lines show in each case the interval $\pm$0.5 dex around $M_d(\lambda)/M_d=1$. The objects are plotted in order of increasing luminosity. `$\vee$' represents a disc mass calculated using a flux density upper limit.}
\label{fig:disk_mass}
\end{figure}

The results of this comparison are shown in Fig.~\ref{fig:disk_mass}, which plots the ratio of monochromatic vs. the `true' disc mass as a function of wavelength for each of the 8 objects. One can see that in most cases the monochromatic masses in the FIR and submm regime are all within a factor of 3 of the `true' values from modeling, which we consider to be the typical uncertainty in disc masses. The deviations are straightforward to explain with scatter in the opacity and in the dust temperature, which are the main sources of uncertainty. Adopting a different temperature in the range 20-30\,K does not significantly change this result. However, we note that using a single temperature for a sample spanning a large range of stellar luminosities may introduce a systematic effect as it tends to underestimate the disc mass at low luminosity and overestimate it at high luminosity. Our sample is skewed toward high luminosity objects, and does not allow us to further explore this effect. We conclude that fluxes at the Herschel wavelengths 160-500$\,\mu$m can be used to estimate robust disc masses for large samples without detailed modeling.

\citet{kim_2009} calculate monochromatic disc masses for 5 of our objects (all TDs), based on 1.3\,mm fluxes. They use almost the same dust opacity ($0.02\,\mathrm{cm}^2 \ \mathrm{g}^{-1}$ at 1.3\,mm), but a significantly higher temperature of $T_c = 50$\,K. These seems unrealistic; using our model, we obtain significantly lower temperatures for the outer regions of the discs. Still, their monochromatic values agree with ours within a factor of three.\footnote{Note that for three out of these five, T21, T25 and T35, \citet{kim_2009} actually use upper flux limits at 1.3\,mm from \citet{henning_1993}, but the resulting disc masses are not listed as upper limits}. For one of our objects, ESO Ha 559, \citet{oloffson_2013} determine a disc mass from radiative transfer modeling and a specified grain size distribution. Their result of 0.005$\,M_{\odot}$ is very similar to ours (0.0055$\,M_{\odot}$). 

\subsection{Disc Masses} 
\label{subsec:disc_mass2}

Having established that Herschel fluxes at $\ge 160\,\mu$m can be utilized to estimate disc masses, we have used Eq.~\ref{eq:m_d} to
calculate monochromatic disc masses for all objects with at least one flux measurement at 160-500$\,\mu$m, excluding
the Class I sources which may be affected by emission from the envelope. The
results are given in Table~\ref{table:dm}. In the last column, we also list the average disc masses for the Herschel
wavelengths; these averages are adopted for the following analysis. Based on the evaluation in Section~\ref{subsec:disc_mass1}, 
the typical uncertainty in these values is a factor of 3. In total, disc masses were obtained for 34 objects, the largest sample of disc masses for the Cha-I region published thus far, about twice as many as in \citet{henning_1993}.

\begin{table*}
\centering
\caption{Monochromatic disc masses derived from the Herschel fluxes and their average for objects in Cha-I.
All values are in $M_{\odot}$. See text for details.}
\begin{tabular}{llcccccc}
\hline
2MASS Coordinate  & Other       & $M_{star}$             &                        & $M_d$ ($M_{\odot}$)  &                       &                       &        \\       
Identifier        & Identifier  & [$M_\odot$]            & 160 [$\mu\mathrm{m}$]  & 250 [$\mu\mathrm{m}$]& 350 [$\mu\mathrm{m}$] & 500 [$\mu\mathrm{m}$] & Average\\ 
\hline
J10563044-7711393 & SY Cha 	 	& 0.5	    & 0.004  & 0.004  & 0.006  & 0.008  & 0.006 \\		    
J10581677-7717170 & SZ Cha	 	& 2.2	    & 0.02   & 0.01   & 0.02   & 0.02	& 0.02  \\
J10590699-7701403 & CR Cha	 	& 1.9	    & 0.01   & 0.01   & 0.02   & 0.02	& 0.02  \\
J11022491-7733357 & CS Cha	 	& 1.0	    & 0.01   & 0.008  & 0.008  & 0.01	& 0.01  \\
J11040909-7627193 & CT Cha	 	& 1.1	    & 0.004  & 0.003  & 0.004  & 0.004  & 0.004 \\
J11061540-7721567 & T21	 	 	& 2.5	    & 0.04   & --    & --    & --    & 0.04  \\
J11062554-7633418 & ESO H$\alpha$ 559	& 0.2	    & 0.002  & 0.001  & 0.002  & --    & 0.002 \\
J11070919-7723049 & Ced110-IRS4	 	& --	    & 0.03   & 0.02   & 0.03   & 0.06	& 0.03  \\
J11071206-7632232 & UZ Cha		& 0.5	    & 0.001  & 0.0007 & --    & --    & 0.0009 \\
J11071622-7723068 & ISO97		& --	    & 0.009  & 0.009  & --    & 0.006  & 0.008  \\
J11071915-7603048 & T25	 	 	& 0.4	    & 0.002  & 0.001  & 0.002  & --    & 0.002  \\
J11072074-7738073 & DI Cha	 	& 2.5	    & 0.008  & --    & --    & --    & 0.008  \\
J11072142-7722117 & B35	 	 	& --	    & 0.002  & --    & --    & 0.008  & 0.005  \\
J11074366-7739411 & FI Cha	 	& 0.6	    & 0.002  & 0.003  & --    & --    & 0.003  \\
J11075792-7738449 & FK Cha	 	& 0.9	    & 0.03   & 0.02   & 0.02   & --    & 0.02  \\
J11080148-7742288 & VW Cha	 	& 0.7	    & 0.003  & --    & --    & --    & 0.003 \\
J11080297-7738425 & ISO126	 	& 0.4	    & 0.009  & 0.008  & 0.003  & --    & 0.007 \\
J11081509-7733531 & Glass I	 	& 2.5	    & 0.02   & 0.01   & 0.01   & 0.01	& 0.01  \\
J11083905-7716042 & T35		 	& 0.7	    & 0.001  & --    & --    & --    & 0.001 \\
J11092266-7634320 & C1-6	 	& 0.5	    & 0.003  & --    & --    & --    & 0.003 \\
J11092379-7623207 & VZ Cha	 	& 0.9	    & 0.002  & 0.002  & 0.003  & 0.006  & 0.003 \\
J11094192-7634584 & C1-25	 	& --	    & 0.01   & --    & 0.001  & 0.02   & 0.01  \\
J11094742-7726290 & B43		 	& 0.3	    & 0.001  & 0.002  & 0.003  & 0.004  & 0.003 \\
J11095340-7634255 & FM Cha	 	& 1.1	    & 0.04   & 0.02   & 0.02   & 0.01	& 0.02  \\
J11095407-7629253 & Sz33	 	& 0.4	    & 0.002  & --    & --    & --    & 0.002 \\
J11095505-7632409 & C1-2	 	& --	    & 0.01   & --    & --    & --    & 0.01  \\
J11095873-7737088 & WX Cha	 	& 0.5	    & 0.001  & --    & --    & --    & 0.001 \\
J11100010-7634578 & WW Cha	 	& 1.1	    & 0.1    & 0.1    & 0.09   & 0.1	& 0.1	\\
J11100369-7633291 & Hn11	 	& 0.7	    & 0.005  & --    & --    & --    & 0.005 \\
J11100704-7629376 & WY Cha	 	& 0.6	    & 0.0006 & --    & --    & --    & 0.0006 \\
J11104959-7717517 & HM27	 	& 0.4	    & 0.002  & --    & --    & --    & 0.002 \\
J11113965-7620152 & XX Cha	 	& 0.4	    & --     & 0.0007 & 0.001  & --    & 0.0009 \\
J11114632-7620092 & CHX 18N	 	& 0.9	    & 0.001  & 0.001  & 0.001  & --    & 0.001 \\
J11122772-7644223 & CV Cha	 	& 2.5	    & 0.007  & 0.004  & 0.003  & 0.004  & 0.004 \\
\hline
\end{tabular}
\label{table:dm}
\end{table*}

\begin{figure}
\centering
\includegraphics[width=0.53\textwidth]{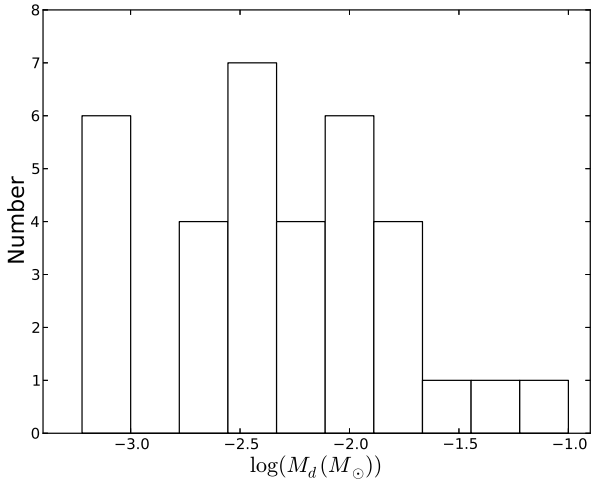}
 \caption{Histogram of the monochromatic disc masses calculated using the Herschel fluxes, see Section~\ref{subsec:disc_mass2}.}
\label{fig:dm_hist}
\end{figure}

This sample can be used to study the overall distribution of disc masses in Cha-I. The histogram of the 34 values
in our sample is shown in Fig.~\ref{fig:dm_hist}. The disc masses range from $10^{-3}$ to $10^{-1}\,M_{\odot}$,
with a median of 0.005$\,M_{\odot}$ for Class IIs. This number is similar to the median disc mass found for Class IIs in the coeval regions Taurus (0.003$\,M_{\odot}$, \citet{andrews_2005}), $\rho$\,Oph (0.005$\,M_{\odot}$,
\citet{andrews_2007} and IC348 (0.002$\,M_{\odot}$, \citet{lee_2011}), although the samples are not 
fully comparable due to different selection effects and methods. We stress again that our sample only covers 
the most luminous discs in Cha-I, i.e. the median value should be interpreted as an upper limit. We also note
that the scatter in disc masses is 2-3 orders of magnitude and thus significantly larger than the uncertainty,
a clear sign of dispersion in the dust content in an approximately coeval population of young discs. 

However, TDs do not have on average lower or higher disc masses than the rest of the sample. The median disc mass for the objects classed in Section~\ref{sec:colour} as TDs is 0.006$\,M_{\odot}$. 

\citet{andrews_2011} studied a sample of 12 TDs from different star-forming regions. They found that their sample of TDs had similar disc masses to the Class IIs in Ophiuchus \citep{andrews_2010}. This agrees with what we have found for Cha-I. In contrast to this, \citet{najita_2007} found that for a sample of 12 TDs from Taurus the median disc mass for TDs was 4 times larger than for the Class IIs in Taurus. All these TD samples are relatively small, and the selection criteria differ. It is therefore not clear whether the difference in the results bears any statistical significance.

Only two objects in our sample have disc masses clearly above the `Minimum Mass Solar Nebula' (MMSN) of around 0.02$\,M_{\odot}$,
which corresponds to the mass of gas and dust required to make the solar system planets \citep{davis_2005}. A handful more
are around this limit. Since we are sampling the most luminous discs in this region and the total number of stars with discs
in Cha-I is around 100 \citep{luhman_2008}, the fraction of objects with MMSN discs is between 2 and 7\%, consistent with
what has been estimated by \citet{greaves_2010} based on the earlier survey by \citet{henning_1993}. Contrasted with the 
abundance of exoplanetary systems with massive planets, our results confirm the `missing mass' problem stated by \citet{greaves_2010} and suggests that the majority of these discs will either never form Jupiter-mass planets or that the formation of larger bodies is already well underway. We note that most of the disc surveys collated by \citet{greaves_2010} use a mm dust opacity similar to the one adopted by us. 

\begin{figure}
\centering
\includegraphics[width=0.53\textwidth]{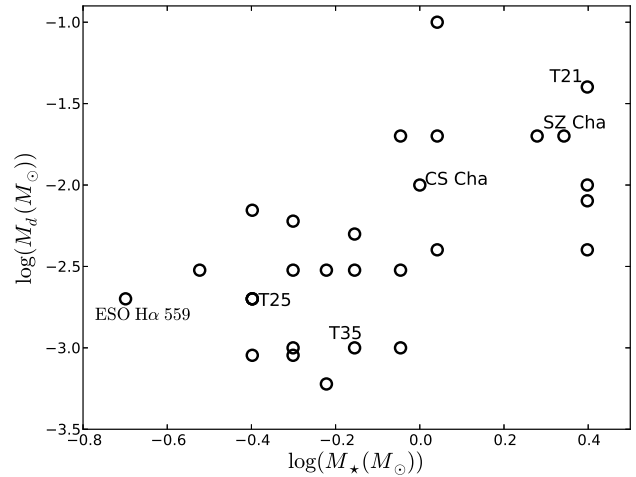}
 \caption{Disc masses in Cha-I plotted against stellar masses: with symbols as in Fig.~\ref{fig:k-12-12-70}. See text for details.}
\label{fig:dm_m}
\end{figure}

We use the Herschel disc masses to probe the relation between disc and stellar mass. For this purpose, we remove 5 objects
without known stellar parameters (4 of them are classified as `flat spectrum'). All remaining sources are either Class II
or TDs. We estimate stellar masses for this sample by comparing the effective temperatures with a 3\,Myr isochrone
from \citet{siess_2000}.\footnote{\tt{http://www.astro.ulb.ac.be/$\sim$siess/WWWTools/Isochrones}} This isochrone was
found to reproduce the trend seen in the HR diagram and thus represents a good approximation 
for our sample. The disc masses are plotted vs. stellar masses in Fig.~\ref{fig:dm_m}.

At first glance, the figure may indicate a trend of larger disc masses towards higher mass objects. However, we found 
no statistically significant correlation between disc masses and stellar masses, using various correlation
tests (linear regression, Rank correlation test). Given the large scatter, the relatively small number of datapoints,
and the relatively small range in masses this is not surprising. Previous studies who have found such trends 
(most notably, the recent paper by \citet{andrews_2013}) covered at least two orders of magnitude in stellar mass,
whereas our sample is limited to one order of magnitude. 

In terms of relative disc masses ($M_d/M_{\mathrm{\star}}$), the median in our sample is 0.5\%, which is consistent with previous studies in Taurus and
$\rho$\,Ophiuchus \citep{mohanty_2013,andrews_2007,andrews_2013}. This indicates that the disc evolution in these
four regions is progressing on similar timescales, without strong environmental difference.

\section{Transition Discs}
\label{sec:tds}

In the following section we briefly discuss the subsample of TDs, i.e. the objects with evidence for some degree of dust clearing in the inner disc which separates them from the Class IIs in Fig.~\ref{fig:k-12-12-70} and Fig.~\ref{fig:col-k-12-70}. We consider here 7 objects with evidence of a deficit of an emission in the NIR/MIR and a rise in the SED between 10 and 20$\,\mu$m.

We use the simple model described in Appendix~\ref{subsec:model} to describe the SEDs of the TDs, specifically, to obtain values for the inner radius of the disc. To create an SED of a TD using this model we merely set the first value of $r$ in Eq.~\ref{eq:temp} to be much larger than $r_0$ and we called this initial value $r_{td}$, whereas for a normal Class II source the first value of $r$ is equal to $r_0$. We varied $r_{td}$ until our model matched the increase in flux density around 10-20$\,\mu$m. Then we varied $T_0, q, \beta, R_d$ and $i$ from their initial fiducial values mentioned in Appendix~\ref{subsec:model} to fit the rest of the SED. We kept within the ranges that are considered to be reasonable for these parameters, which are given in Appendix~\ref{subsec:model}. Note that this model does not allow us to fit the silicate feature or any other excess emission bluewards of the steep increase in flux density.

\begin{figure*}

\centerline{\includegraphics[width=0.3\textwidth]{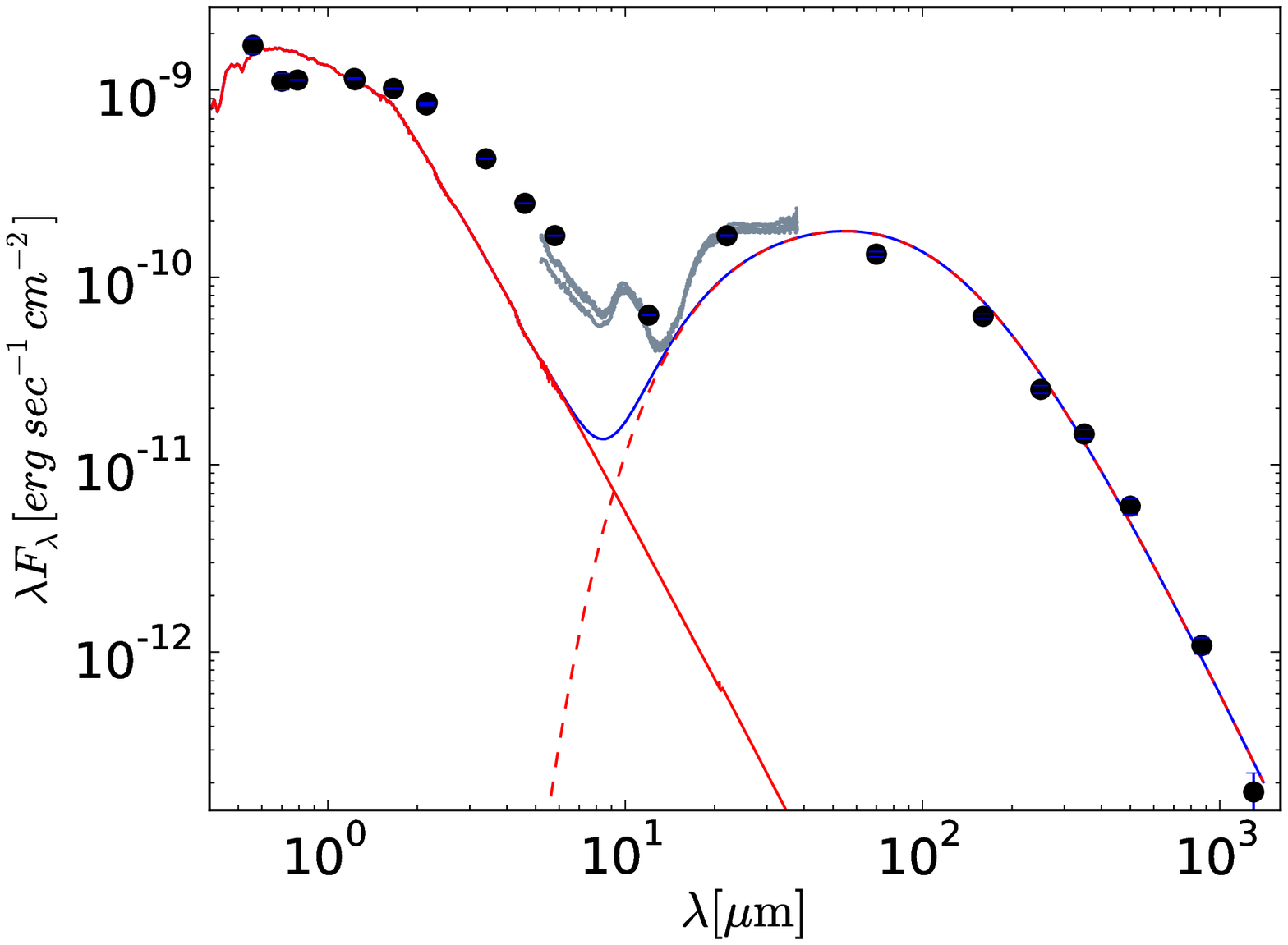}\qquad
\includegraphics[width=0.3\textwidth]{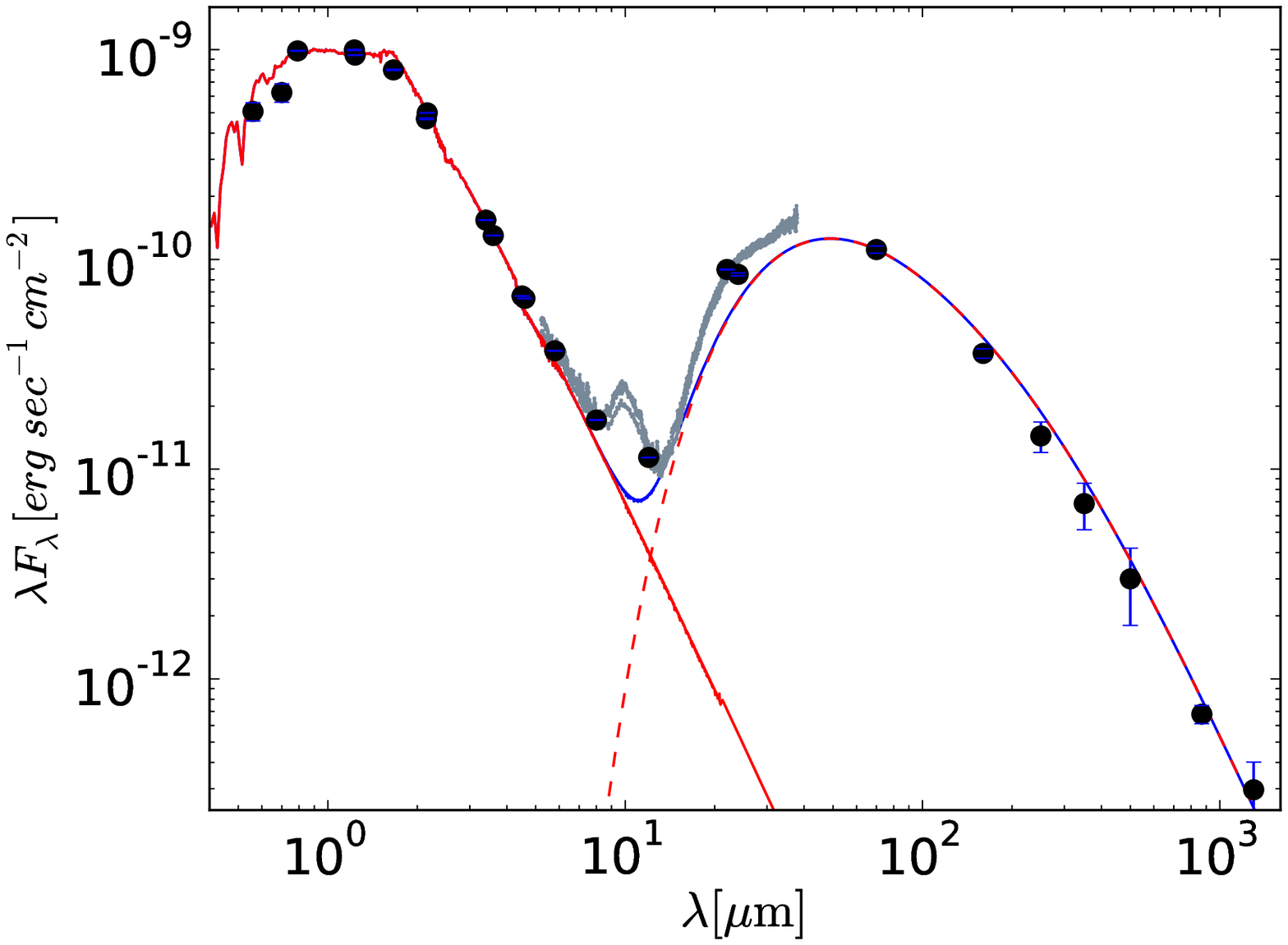}\qquad
\includegraphics[width=0.3\textwidth]{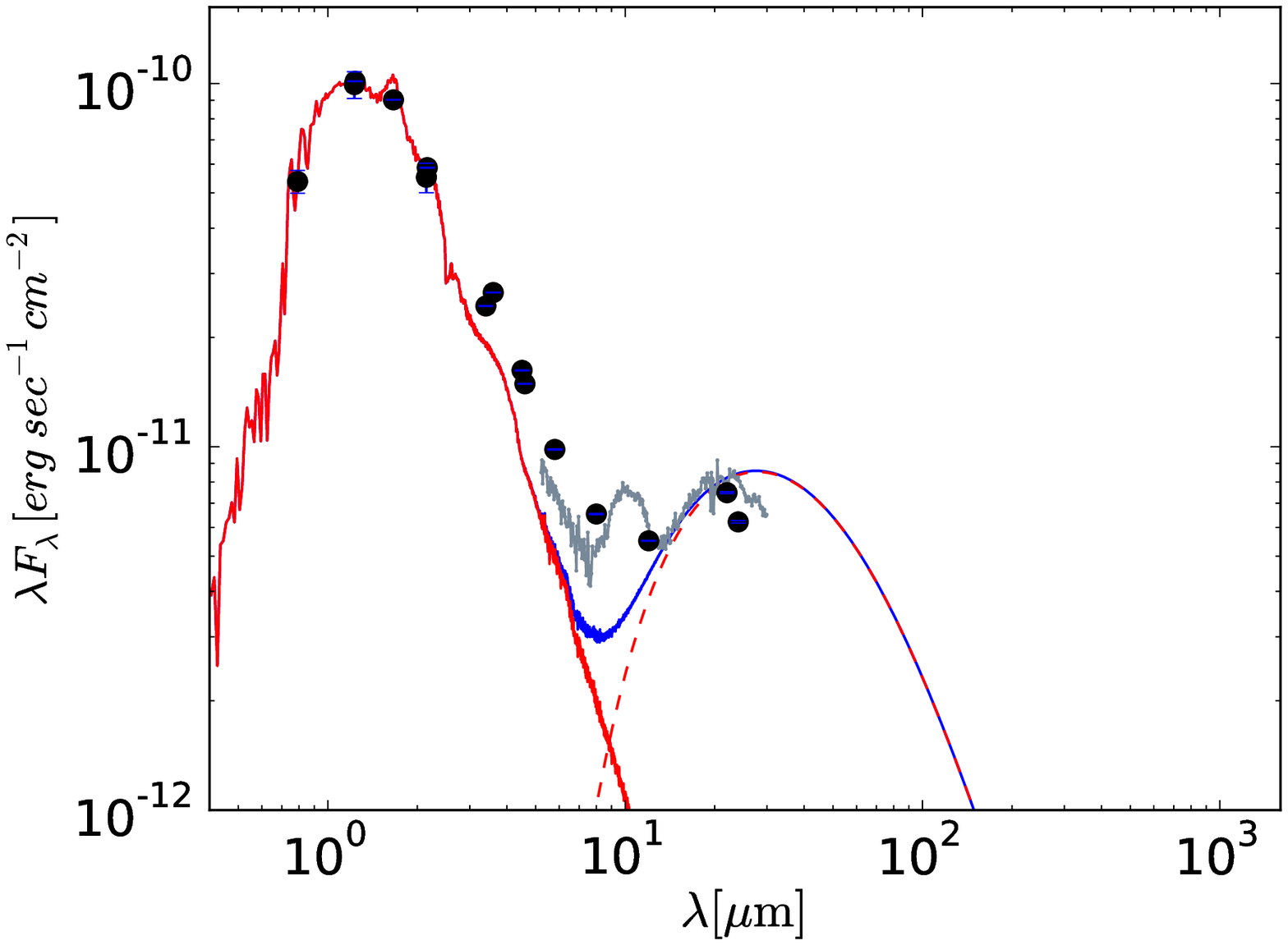}}
\centerline{(a) SZ Cha \hspace{0.25\textwidth} (b) CS Cha \hspace{0.25\textwidth} (c) ISO 52 }

\centerline{\includegraphics[width=0.3\textwidth]{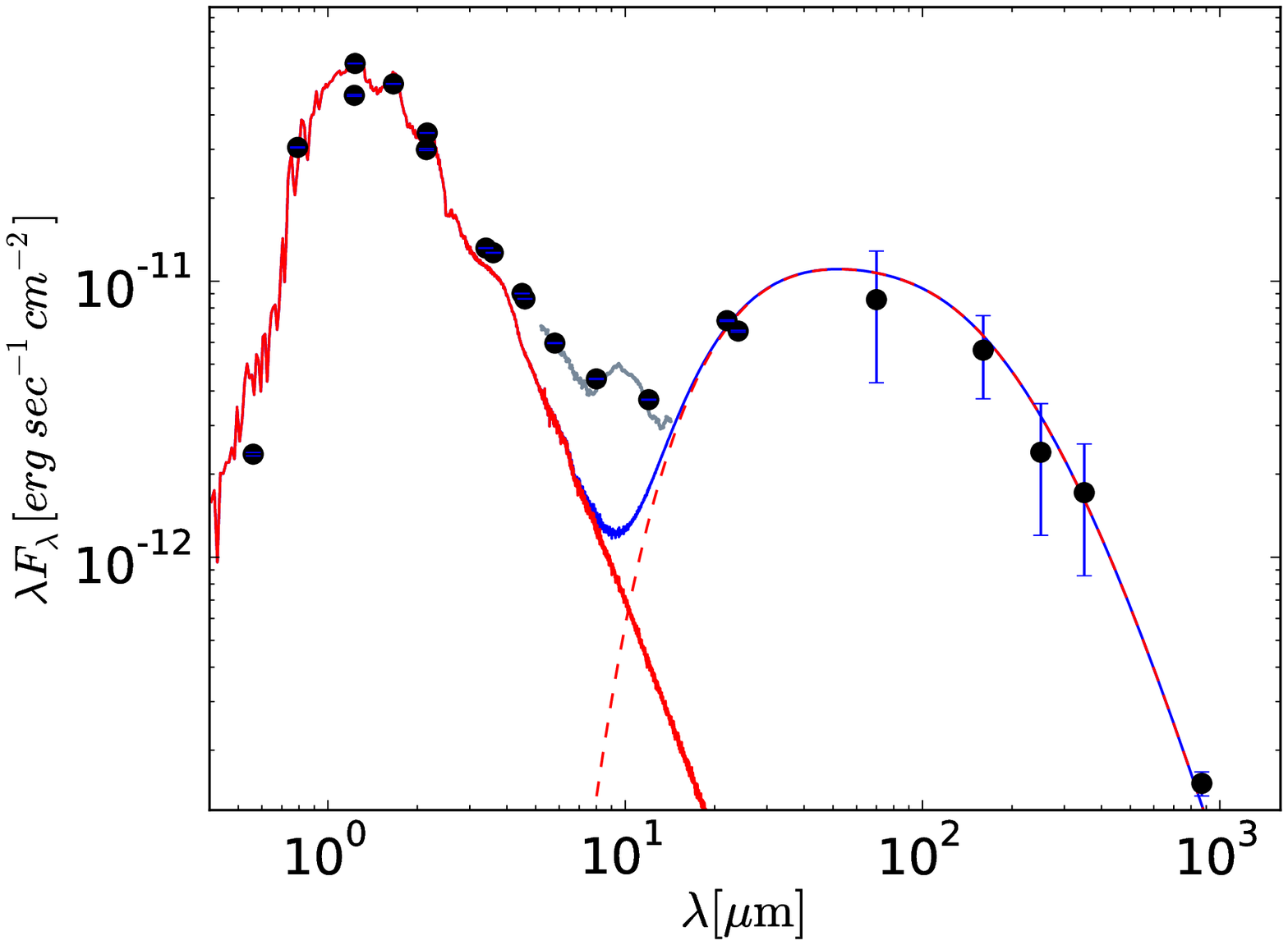}\qquad
\includegraphics[width=0.3\textwidth]{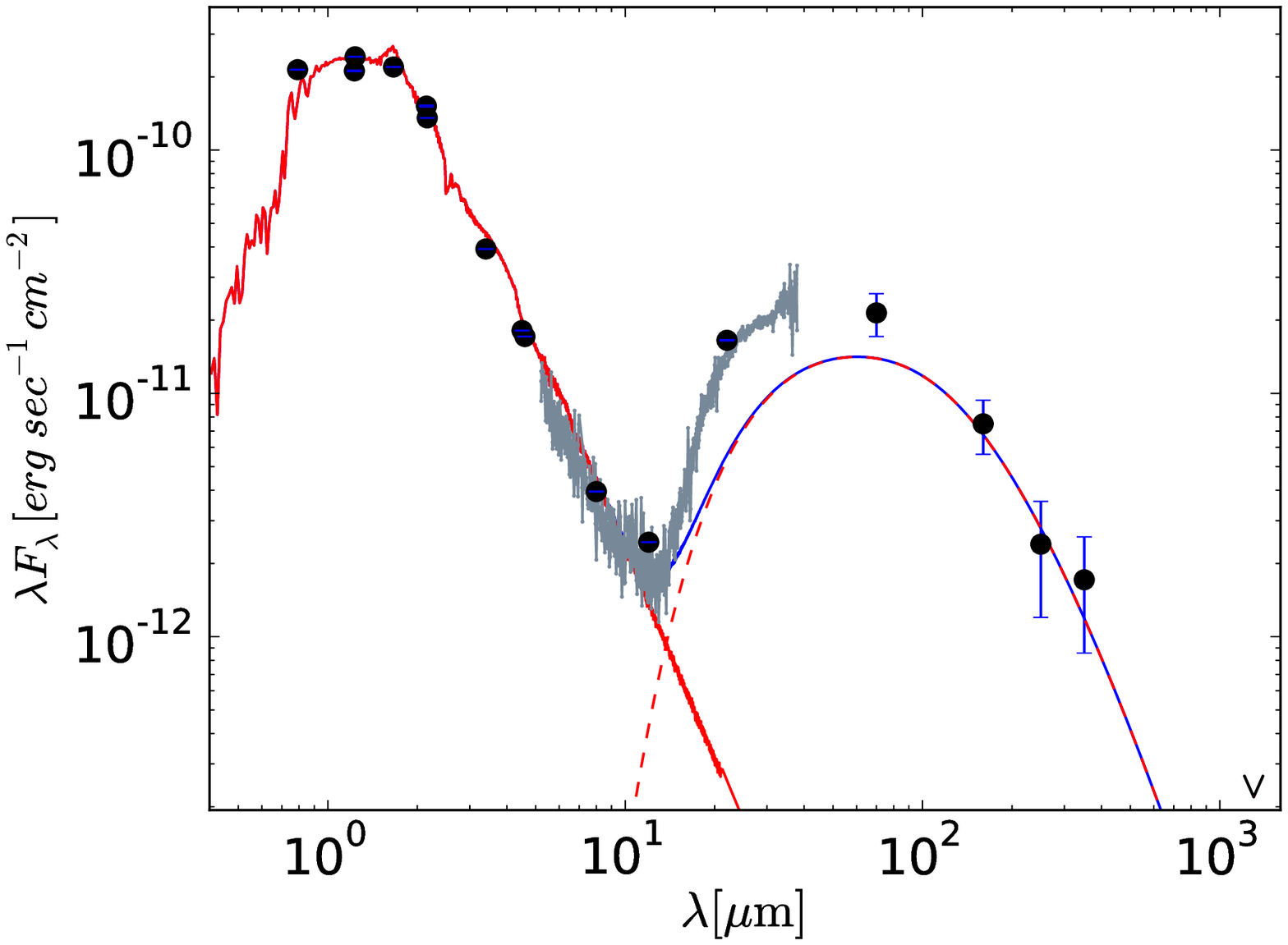}\qquad
\includegraphics[width=0.3\textwidth]{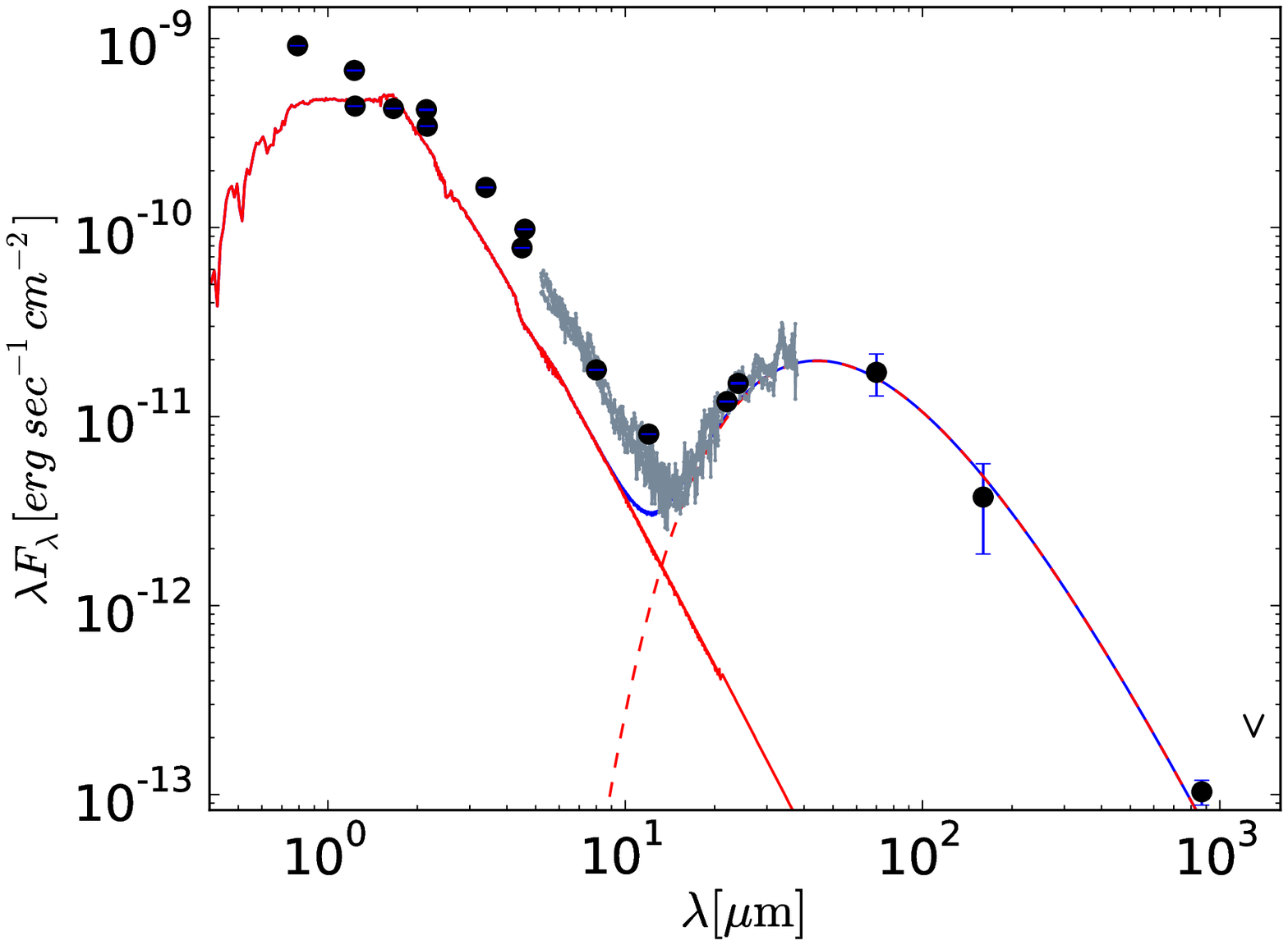}}
\centerline{(d) ESO H$\alpha$ 559 \hspace{0.25\textwidth} (e) T25 \hspace{0.25\textwidth} (f) T35 }

\caption{ SEDs of TDs fitted using the simple model (see Appendix~\ref{subsec:model}). The red line is the model atmosphere, the dotted red line is the simple model fit and the blue line is combination of both. \label{fig:seds_td}} 
\end{figure*}

The SEDs with the matching model are shown in Fig.~\ref{fig:seds_td}; their parameters are listed in Table~\ref{table:td_values}. Note that the SED for T21 is instead given in Fig.~\ref{fig:seds}\,(c). In particular, we found values for $r_{td}$ ranging from 0.185\,-\,13.3\,AU. The SEDs reveal a large diversity of features which reflects the variety of physical conditions and geometry of the TDs. Four objects out of 7, Figs.~\ref{fig:seds_td}\,(c),(d),(f) and Fig.~\ref{fig:seds}\,(c) are well matched by our simple model from 10\,$\mu$m onwards. For the other 3 objects, however, Figs.~\ref{fig:seds_td}\,(a),(b),(e), the simple model underestimates the flux density between 10\,-\,70\,$\mu$m. For these objects which are poorly matched by our model the assumptions made in our approach are probably not suitable.  

\citet{kim_2009} model 5 out of the 7 TDs that we have investigated. The inner radii that they obtained are also shown in Table~\ref{table:td_values}; they range from 8.1\,-\,146.7\,AU and are about an order of magnitude larger than ours. This discrepancy is a result of the different assumptions made when modeling the SED. In our simple model, we consider the dusty disc to be truncated, without any further modification at the inner edge of the disc. We assume that the inner edge of the disc is not directly irradiated by the star. On the other hand, \citet{kim_2009} (in line with other authors, e.g. \citet{dalessio_2005,espaillat_2007c,brown_2007,espaillat_2007d}) use more complex models which assume that the inner edge of the outer optically thick disc is directly exposed to the stellar radiation. As a consequence, a so-called wall will develop. Since the rise in the SED essentially constrains the maximum temperature of the optically thick disc, assuming that the edge of the disc is directly exposed to the stellar radiation makes it hotter; therefore the same temperature is reached further out from the star, i.e. $r_{td}$ is increased. Thus, the difference between our results and the ones from \citet{kim_2009} is as expected.

It is clear that neither of these models is entirely correct as some amount of gas and dust inside the inner cavity is observed in most TDs. Most of the TDs in our sample show some level of excess emission in the NIR and MIR bluewards of the rise in the SED (see Fig. \ref{fig:seds_td}), indicating the presence of dust in the cavity. Moreover, two of the TDs have quite high accretion rates of $\sim 10^{-8}\,M_\odot \ \mathrm{yr^{-1}}$ \citep{antoniucci_2011}, see Table \ref{table:entire_sample}. This means that gas is also present near the central object. The effect of this material on the location and development of an inner wall needs to be addressed using more realistic 2D models. Here we are merely stressing the very large uncertainty that any simple assumption will have on the determination of $r_{td}$ from the SED.

Resolved imaging of the gaps in the discs is only available for one of the objects discussed here. \citet{cieza_2013} use Sparse Aperature Masking (SAM) to investigate T35. They find an inner disc, a gap and then an outer disc, with its inner edge at $\sim$8.3\,AU. Using submm interferometry, \citet{cieza_2013} find 8.1\,AU cavity radius for the same object. These results are approximately an order of magnitude larger than our $r_{td}$ inferred from the SED (0.95\,AU). More indirect constraints can be gained by looking for stellar companions. If a companion is detected within the disc, the cavity radius has to be at least as large as the distance between object and companion. While all our TDs have been observed with resolution of $\sim 0\farcs1$ (corresponding to $\sim 15$\,AU), either with Adaptive Optics at the VLT \citep{lafreniere_2008} or with the HST \citep{robberto_2012}, only two of them have a confirmed companion. SZ Cha has a wide companion at 800\,AU, i.e. presumably outside the disc. T21 has a companion at $\sim 22$\,AU, about twice as large as our estimate for $r_{td}$. Thus, for two of our objects, T21 and T35, our values for $r_{td}$ are too small, indicating the necessity of considering more complex models, possibly the inclusion of a wall-like feature. Note that these two are well approximated by our simple model (see Fig. \ref{fig:seds_td}).

Based on the observational parameters, we can speculate about the possible origins of the cavities in the TDs in Cha-I. In general, photoevaporation models for TDs predict low accretion rates, small inner hole sizes and low disc masses \citep{armitage_2007,Owen_2011}.The known accretion rates for our sample are most likely too high for photoevaporation scenarios. In addition, most of the disc masses estimated for our sample are relatively large compared with the predictions for photoevaporation. Thus, alternative scenarios need to be considered. The presence of close-in companions or massive planets inside the gaps is a plausible option for the majority of these TDs. 

\begin{table*}
 \centering
  \caption{Derived properties of TDs in Cha I using the simple model from Appendix~\ref{subsec:model}}
  \begin{tabular}{@{}lllllllllll@{}}
  \hline
   
  2MASS  & Name & $q$ &  $\beta$  & $r_{td}$ & $r_{\mathrm{Kim}}^{\ a} $  & $\mathrm{T}(r_{td})$ & $\mathrm{T}(r_{\mathrm{Kim}})^{ \ b}$ & $R_d$ & $i$ & $M_d$\\
   Coordinates & &   &   &  [AU] &  [AU] & [K] & [K] & [AU] & & $[M_\odot]$\\
 \hline
  J10581677$-$7717170 & SZ Cha 		   & 0.49  & 0.6  & 1.10 & 29.5 & 195 & 90  & 100 & 60$^\circ$ & 0.015 \\
  J11022491$-$7733357 & CS Cha  	   & 0.54  & 0.3  & 2.2  & 41.1 & 127 & 62  & 150 & 0$^\circ$ & 0.02 \\ 
  J11044258$-$7741571 & ISO 52		   & 0.59  & 0.8  & 0.185 & -	& 216 & -   & 100 & 50$^\circ$ & $<0.0003$ \\
  J11061540$-$7721567 & T21/Ced 110 IRS 2  & 0.58  & 1.0  & 13.3 & 146.7& 56  & 66  & 150 & 0$^\circ$  & 0.03 \\
  J11062554$-$7633418 & ESO H$\alpha$ 559  & 0.55  & 0.9  & 0.4  & -	& 193 & -   & 100 & 55$^\circ$ & 0.0055 \\ 
  J11071915$-$7603048 & T25 		   & 0.57  & 1.2  & 1.0  & 8.1  & 132 & 90  & 150 & 50$^\circ$ & 0.002 \\ 
  J11083905$-$7716042 & T35/FL Cha  	   & 0.6   & 0.3  & 0.95 & 15.2 & 137 & 80  & 100 & 40$^\circ$ & 0.002 \\

\hline

\end{tabular}

$^a \ r_{\mathrm{Kim}} $ is $r_{td}$ as calculated by \citet{kim_2009}, $^b \ \mathrm{T}(r_{\mathrm{Kim}}) $ is the temperature at $r_{td}$ calculated by \citet{kim_2009} \\
\label{table:td_values}

\end{table*}

\section{Summary and Conclusions}
\label{sec:conclusion}
We have carried out a multi-wavelength study of circumstellar discs around 44 YSOs in the Cha-I star-forming region. In particular we made use of the recent release of Herschel data for Cha-I. For a sample of 8 test cases (all of which have either a sub-mm or mm flux measurement) we examined whether Herschel fluxes at a single wavelength could be used to derive disk masses (monochromatic masses). We found that Herschel fluxes at 160-500$\,\mu$m can be used to derive robust estimates of the disc mass. In most cases the monochromatic masses in the FIR and submm regime are within a factor of 3 of the `true' values from modeling. The deviations can be explained by a scatter in the opacity and in the dust temperature, which are the main sources of uncertainty.

In total, disc masses were obtained for 34 objects, the largest sample of disc masses published thus far for the Cha-I region, about twice as many as in \citet{henning_1993}. The median disc mass is 0.005$\,M_{\odot}$ for Class IIs, corresponding to on average 0.5\% of the stellar mass, but with a scatter of 2-3 orders of magnitude. The median disc mass for TDs is 0.006$\,M_{\odot}$. The fraction of objects in Cha-I with at least the `minimum mass solar nebula' is only 2-7\%. These numbers are consistent with previously published results in other regions (Taurus, IC348, $\rho$\,Oph). 

Diagrams of spectral slopes, making use of Herschel data, show the effect of specific evolutionary processes in circumstellar discs. We found that in our sample of objects the Class IIs show a wide scatter in the $\alpha_{K-70}$ spectral slope. We ascertained using a 2-layer disc model that this scatter can be explained as a consequence of dust settling. We found that there is a similar scatter in the $\alpha_{K-70}$ spectral slope for TDs. We identify a continuous trend from Class II to TDs which can be accounted for by assuming inner disc clearing. We discuss the implications of simple assumptions for the determination of the inner gap radius from SED fitting. The development of an inner wall needs to be addressed using more realistic 2D models.

Including Herschel fluxes in this type of analysis highlights the diversity of TDs. We find that TDs do not show any significant difference compared with Class IIs in terms of far-infrared luminosity, disc mass, or amount of dust settling. Taken together, this indicates that inner dust clearing occurs independently from other evolutionary processes in the discs.

\section*{Acknowledgments}
We would like to thank the anonymous referee for helpful comments which improved the manuscript.

\appendix

\section{List of properties for objects}

In Table \ref{table:entire_sample} we list the entire sample considered in this paper, including stellar parameters, evolutionary classes, and accretion rates collected from the literature.

\begin{table*}
 \centering
  \caption{Derived properties for our 44 objects in Cha I from \citet{luhman_2007} and accretion rates from \citet{antoniucci_2011,costigan_2012}. }
  \begin{tabular}{@{}lllllllll@{}}
  \hline
Coordinate	  &  Other 		     &   Evol.  & Spectral   & $T_{\mathrm{eff}}$ & $A_J$  & $L_{star}$  & $\dot{M}^{\ a}$ & $\dot{M}^{\ b}$    \\
identifier	  & identifier 		     &   class  &  Type      &  [ K ]		  & -	   & $[L_\odot]$   & [$\dot{M}_\odot \ \mathrm{yr^{-1}}]$ & [$\dot{M}_\odot \ \mathrm{yr^{-1}}]$ \\
\hline				     
J10563044-7711393 & SY Cha, T4, Sz3	     &   II	&	M0.5 &  3778 &  0.23 &    0.57      &  -	    \\      
J10581677-7717170 & SZ Cha, T6  	     &   II/TD  &	  K0 &  5250 &  0.42 &     1.9      &  -	    \\      
J10590699-7701404 & CR Cha, T8, Sz6	     &   II	&	  K2 &  4900 &  0.00 &     2.4      &  -	    \\      
J11022491-7733357 & CS Cha, T11 	     &   TD	&	  K6 &  4205 &  0.07 &     1.2      & $1.96 \times 10^{-08 }$	    \\      
J11040909-7627193 & CT Cha, T14 	     &   II	&	  K5 &  4350 &  0.45 &    0.95      & $1.42 \times 10^{-08}$ \\     
J11042275-7718080 & HH48, T14A  	     &   I	&	  K7 &  4060 &  0.45 &   0.013      &  -	    \\      
J11044258-7741571 & ISO52		     &   II	&	  M4 &  3270 &  0.36 &   0.093      & $1.17 \times 10^{-09}$	    \\      
J11061540-7721567 & Ced110-IRS2, T21	     &   III	&	  G5 &  5770 &  0.92 &      16      &  -	    \\      
J11062554-7633418 & ESO Ha 559  	     &   II	&      M5.25 &  3091 &  1.01 &   0.052      &  -	    \\      
J11063460-7723340 & Cha MMS-1		     &   0	&  -	     &   -   &   -   &      -	    &  -	    \\
J11064658-7722325 & Ced110-IRS4 	     &   I	&  -	     &   -   &   -   &      -	    &  -	    \\
J11070919-7723049 & Ced110-IRS6 	     &   Flat	&  -	     &   -   &   -   &     -	    &  -	    \\
J11071206-7632232 & UZ Cha, T24 	     &   II	&	M0.5 &  3778 &  0.63 &    0.36      &  -	    \\      
J11071622-7723068 & ISO97		     &   Flat	&  -	     &   -   &   -   &     -	    &  -	    \\
J11071915-7603048 & T25, Sz18		     &   II	&	M2.5 &  3488 &  0.45 &    0.24      &  -	    \\      
J11072074-7738073 & DI Cha, T26 	     &   II	&	  G2 &  5860 &  0.75 &      12      & $5.98 \times 10^{-08 }$ &  $6.9 \times 10^{-8 }$  \\      
J11072142-7722117 & B35 		     &   Flat	&  -	     &   -   &   -   &     -	    &  -	    \\
J11074366-7739411 & FI Cha		     &   II	&	  M0 &  3850 &  1.35 &     1.4      &  -	    \\      
J11075730-7717262 & CHXR30b		     &   II	&      M1.25 &  3669 &  3.16 &    0.22      & $7.72 \times 10^{-09 }$	    \\      
J11075792-7738449 & FK Cha, HM16	     &   Flat	&	  K6 &  4205 &  1.24 &     2.4      &  -	    \\      
J11080148-7742288 & VW Cha, T31 	     &   II	&	  K8 &  3955 &  0.72 &     3.0      & $1.67 \times 10^{-07 }$ &  $8.71 \times 10^{-8 }$ \\      
J11080297-7738425 & ISO126		     &   II	&      M1.25 &  3669 &  1.35 &    0.33      & & $2.23 \times 10^{-9 }$		    \\      
J11081509-7733531 & Glass I, T33A	     &   Flat	&	  G7 &  5630 &  0.85 &     2.8      &  -	    \\      
J11083896-7743513 & IRN 		     &   I	&  -	     &   -   &   -   &     -	    &  -	    \\
J11083905-7716042 & T35, ISO151 	     &   II	&	  K8 &  3955 &  1.31 &    0.53      &  -	    \\      
J11085464-7702129 & VY Cha, T38, Sz29	     &   II	&	M0.5 &  3778 &  0.90 &    0.34      & $7.13 \times 10^{-09 }$	    \\      
J11091812-7630292 & CHXR79, Hn9 	     &   II	&      M1.25 &  3669 &  1.92 &    0.55      & $1.58 \times 10^{-08 }$  \\    
J11092266-7634320 & C1-6, P30		     &   II	&      M1.25 &  3669 &  3.27 &    0.80      & $4.74 \times 10^{-08 }$	    \\      
J11092379-7623207 & VZ Cha, T40 	     &   II	&	  K6 &  4205 &  0.56 &    0.53      &  -	    \\      
J11092855-7633281 & ISO192		     &   I	&  -	     &   -   &    -  &     -	    &  -	    \\
J11094192-7634584 & C1-25, ISO199	     &   II	&  -	     &   -   &    -  &     -	    &  -	    \\
J11094742-7726290 & B43, ISO207 	     &   II	&      M3.25 &  3379 &  2.26 &    0.22      & $2.36 \times 10^{-08 }$ & $4.07 \times 10^{-9 }$ \\      
J11095340-7634255 & FM Cha, HM23	     &   II	&	  K5 &  4350 &  1.47 &     3.0      & $1.42 \times 10^{-07 }$	    \\      
J11095407-7629253 & Sz33, T43		     &   II	&	  M2 &  3560 &  1.47 &    0.48      & $1.63 \times 10^{-08 }$	    \\      
J11095505-7632409 & C1-2, ISO226	     &   Flat	&  -	     &   -   &   -   &     -	    &  -	    \\
J11095873-7737088 & WX Cha, T45 	     &   II	&      M1.25 &  3669 &  0.56 &    0.84      & &  $1.51 \times 10^{-8 }$	    \\
J11100010-7634578 & WW Cha, T44 	     &   II	&	  K5 &  4350 &  1.35 &     5.5      &  -	    \\      
J11100369-7633291 & Hn11, ISO232	     &   II	&	  K8 &  3955 &  2.14 &    0.66      &  -	    \\      
J11100704-7629376 & WY Cha, T46		     &   II	&	  M0 &  3850 &  1.13 &     1.4      &  -	    \\      
J11103801-7732399 & CHXR47		     &   II	&	  K3 &  4730 &  1.44 &     2.6      &		    \\      
J11104959-7717517 & HM27, T47		     &   II	&	  M2 &  3560 &  1.17 &    0.42      & $1.41 \times 10^{-08 }$	    \\      
J11113965-7620152 & XX Cha, T49, Sz39	     &   II	&	  M2 &  3560 &  0.34 &    0.37      & $1.55 \times 10^{-08 }$	    \\      
J11114632-7620092 & CHX 18N, Cam1-103	     &   II	&	  K6 &  4205 &  0.20 &     1.3      &  -	    \\      
J11122772-7644223 & CV Cha, T52 	     &   II	&	  G9 &  5410 &  0.42 &     5.0      &  -	    \\      
\hline
\end{tabular}

$^a$ \citet{antoniucci_2011}, $^b$ \citet{costigan_2012} \\
 \label{table:entire_sample}
  \end{table*}

\section{Comments on peculiar objects}
\label{sec:outliers}

As already mentioned in Section~\ref{subsec:discuss_seds} we exclude T54 from the sample as it suffers from extended source contamination \citep{matra_2012}.

\subsection{Candidate edge-on discs}
\label{subsec:edge-on}
\begin{figure*}

\centerline{\includegraphics[width=0.35\textwidth]{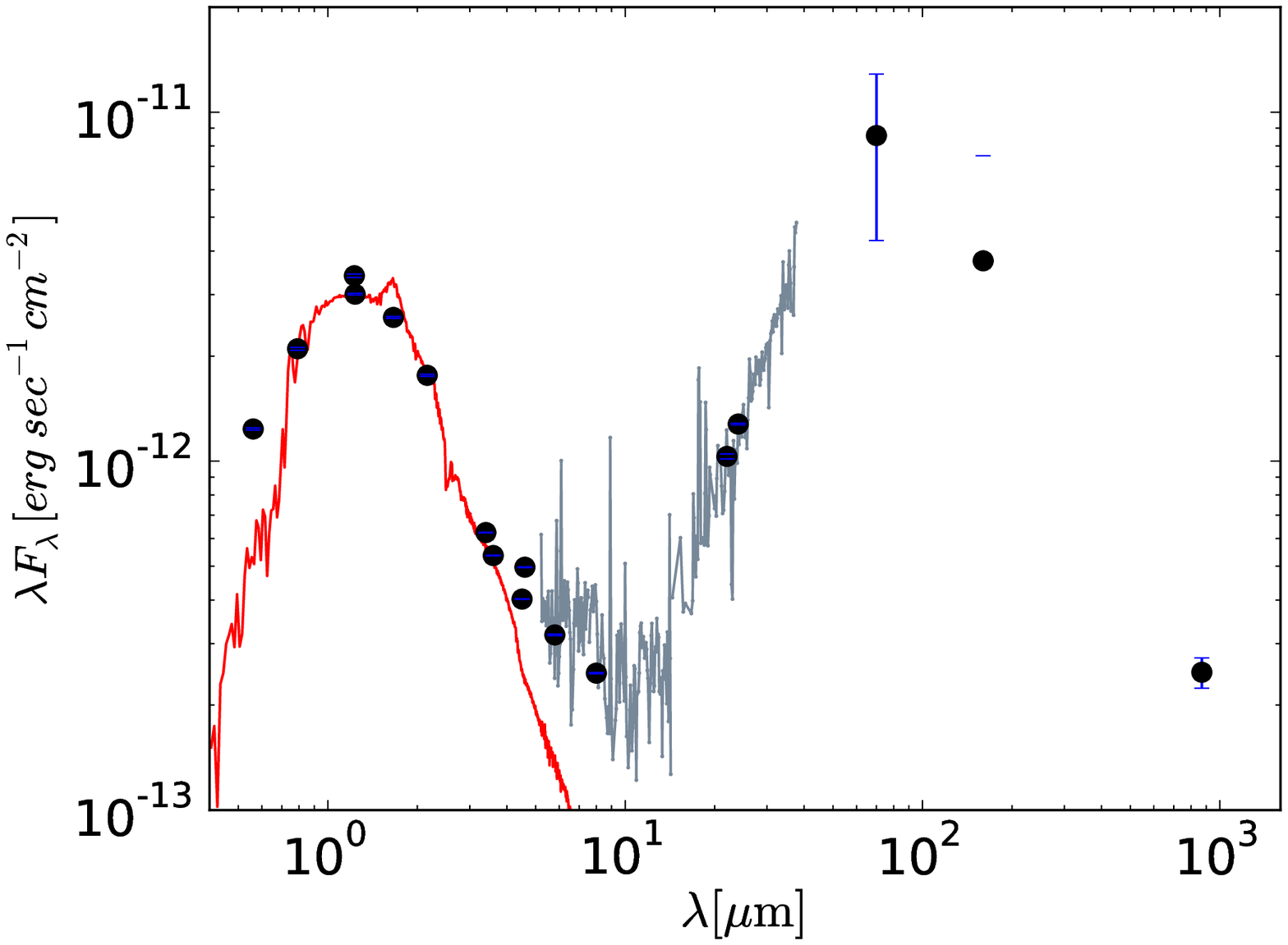}\qquad
\includegraphics[width=0.35\textwidth]{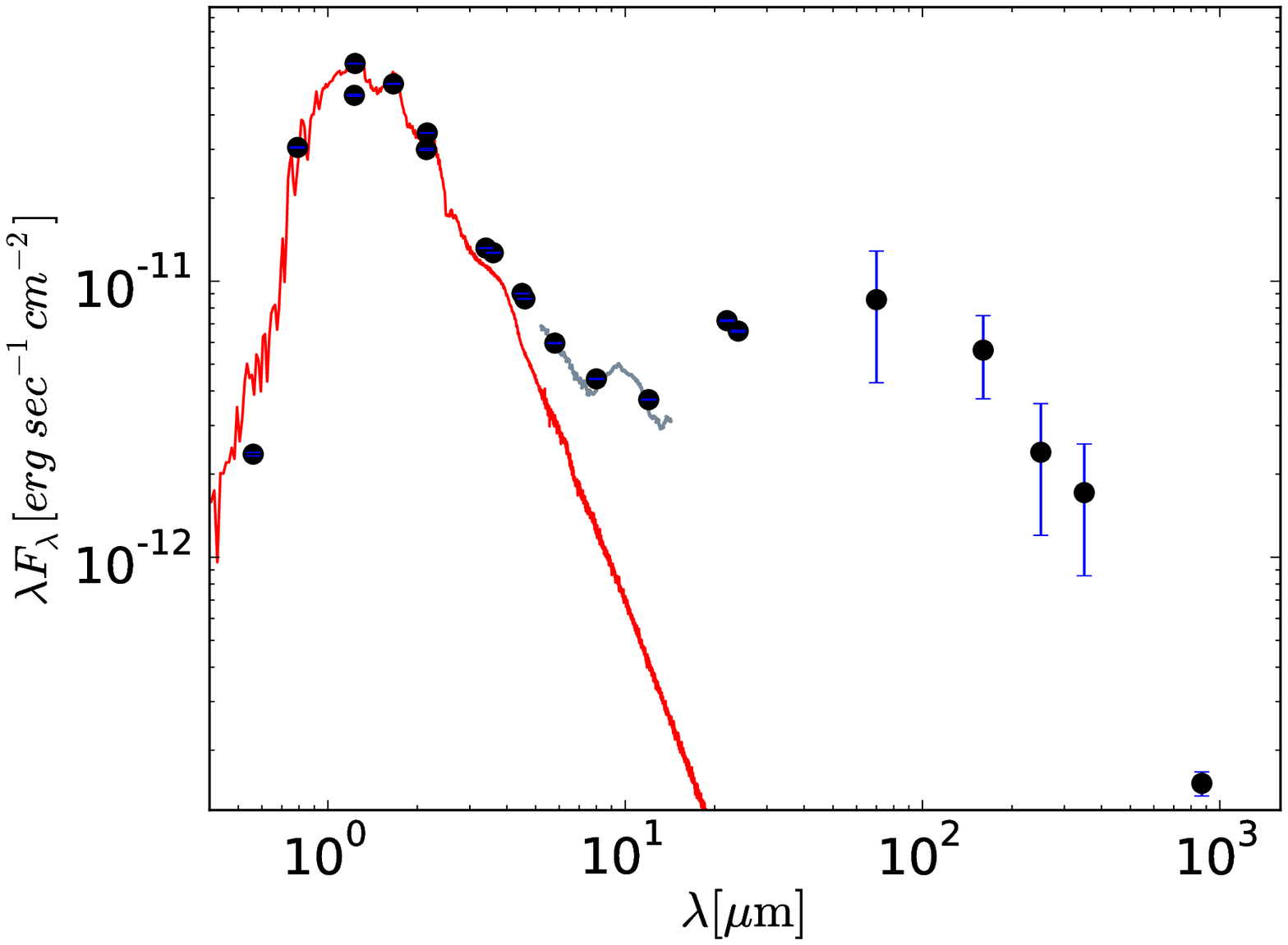}\qquad
\includegraphics[width=0.35\textwidth]{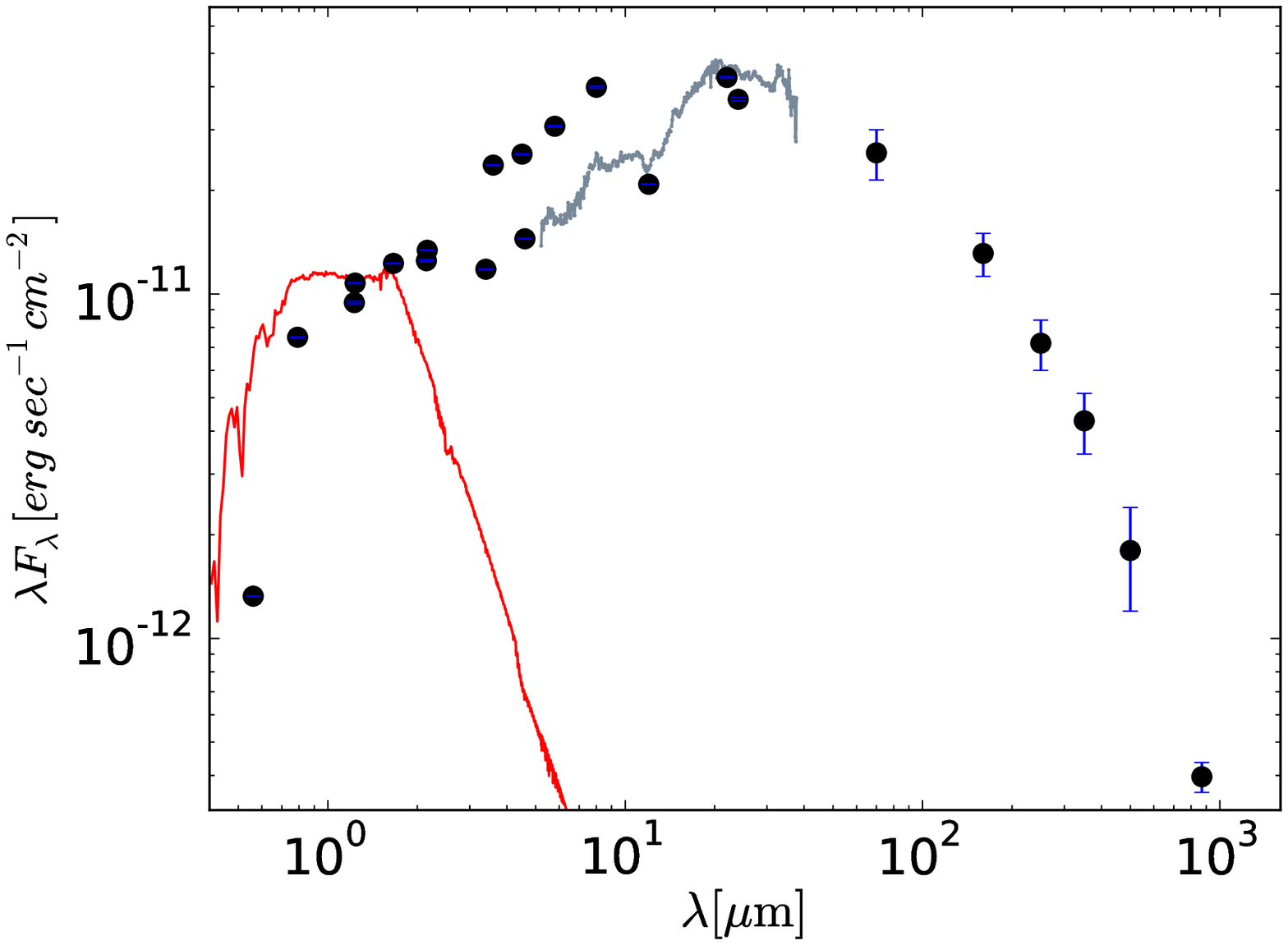}}
\centerline{(a) ESO H$\alpha$ 569\hspace{0.25\textwidth} (b) ESO H$\alpha$ 559 \hspace{0.25\textwidth} (c) HH48 }

\caption{SEDs of outliers in HR diagram and potential edge-on discs.\label{fig:outliers}}
\end{figure*}

We discuss briefly 3 objects that may be edge-on discs. ESO H$\alpha$ 569 and ESO H$\alpha$ 559 appear to be TDs, based on their SEDs. We are interested in the characteristics of TDs, hence it is important for us to clarify whether or not we consider them to be TDs or edge-on discs. Lastly, HH48/T14A is also potentially an edge-on disc. It is class I, from examination of the SED, so it does not affect our TD sample. The SEDs for these objects are given in Fig.~\ref{fig:outliers}.

\subsubsection{ESO H$\alpha$ 569} \label{subsec:569}

\citet{robberto_2012} found ESO H$\alpha $ 569 to be a nearly edge-on disc with an angle of inclination of $87.1^\circ$ using the online SED fitting tool from \citet{robitaille_2006}. We found a best fit of $81.4^\circ$ using the online SED fitting tool, even when we included the {\it Herschel} data which were not available before. Including the LABOCA data point gave a best fit of $18.2^\circ$. \citet{luhman_2007} state that the absence of a detection of this star in the X-ray images from \citet{feigelson_2004} can be used to constrain the amount of extinction towards the star, indicating an extinction of $A_K \gtrsim 60 $ towards this star. This high extinction and the underluminosity of the object in the HR diagram, see Fig.~\ref{fig:hr}, support the conclusion that ESO H$\alpha $ 569 is an edge-on disc. This object is not mentioned in \citet{ribas_2013} but is included in \citet{winston_2012}. In the recent study by \citet{oloffson_2013} it was also concluded that this object is a likely edge-on disc, based on modeling the infrared SED. 

\subsubsection{ESO H$\alpha$ 559}

There are two reasons why this object has previously been classed as edge-on disc.
First, ESO H$\alpha$ 559 was found by \citet{robberto_2012} to be a nearly edge-on disc using the online SED fitting tool from \citet{robitaille_2006}. We attempted to reproduce the results by \citet{robberto_2012} using the same tool, but could not fit the observed {\it Herschel} flux densities for ESO H$\alpha$ 559.

Second, \citet{ribas_2013} state that the underluminosity of ESO H$\alpha$ 559 with respect to its spectral type also supports this scenario. But ESO H$\alpha$ 559 is not underluminous in the HR diagram, see Fig.~\ref{fig:hr}, and so we see no reason to classify it as a nearly edge-on disc. 

\subsubsection{HH48/T14A}

HH48 is a class I source which is underluminous in the HR diagram. \citet{luhman_2004} suggests that HH48 is underluminous because it is an edge-on disc. There is no further information available to suggest that this is indeed the case. In the absence of further information the question remains open. The SED of HH48/T14A is class I and there is no evicence for inner disk clearing or dust settling, i.e. for the purposes of this work it is immaterial whether the object is edge-on or not.

\subsubsection{ISO 237 and Hn10e}

\begin{figure*}

\centerline{\includegraphics[width=0.35\textwidth]{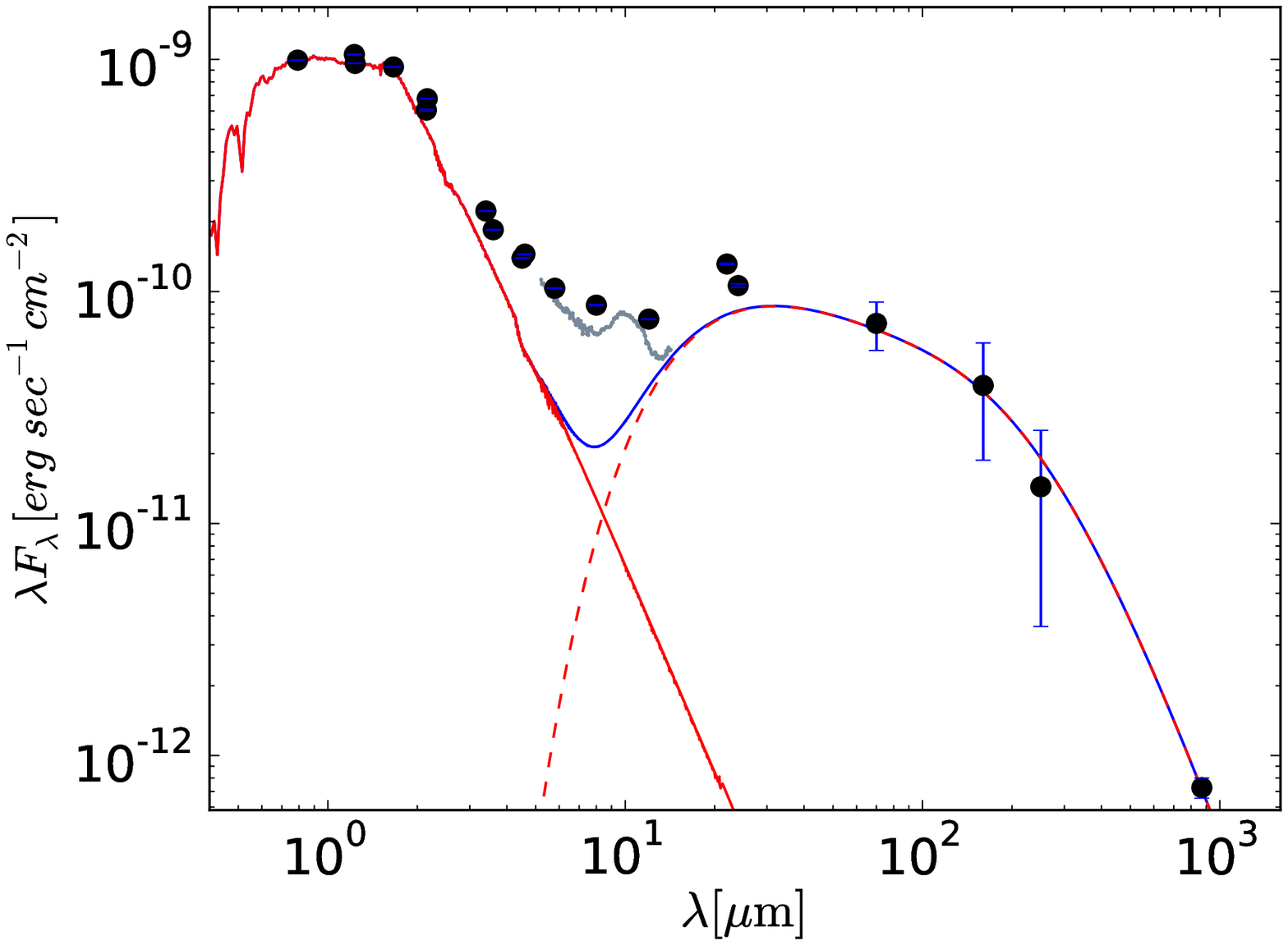}\qquad
\includegraphics[width=0.35\textwidth]{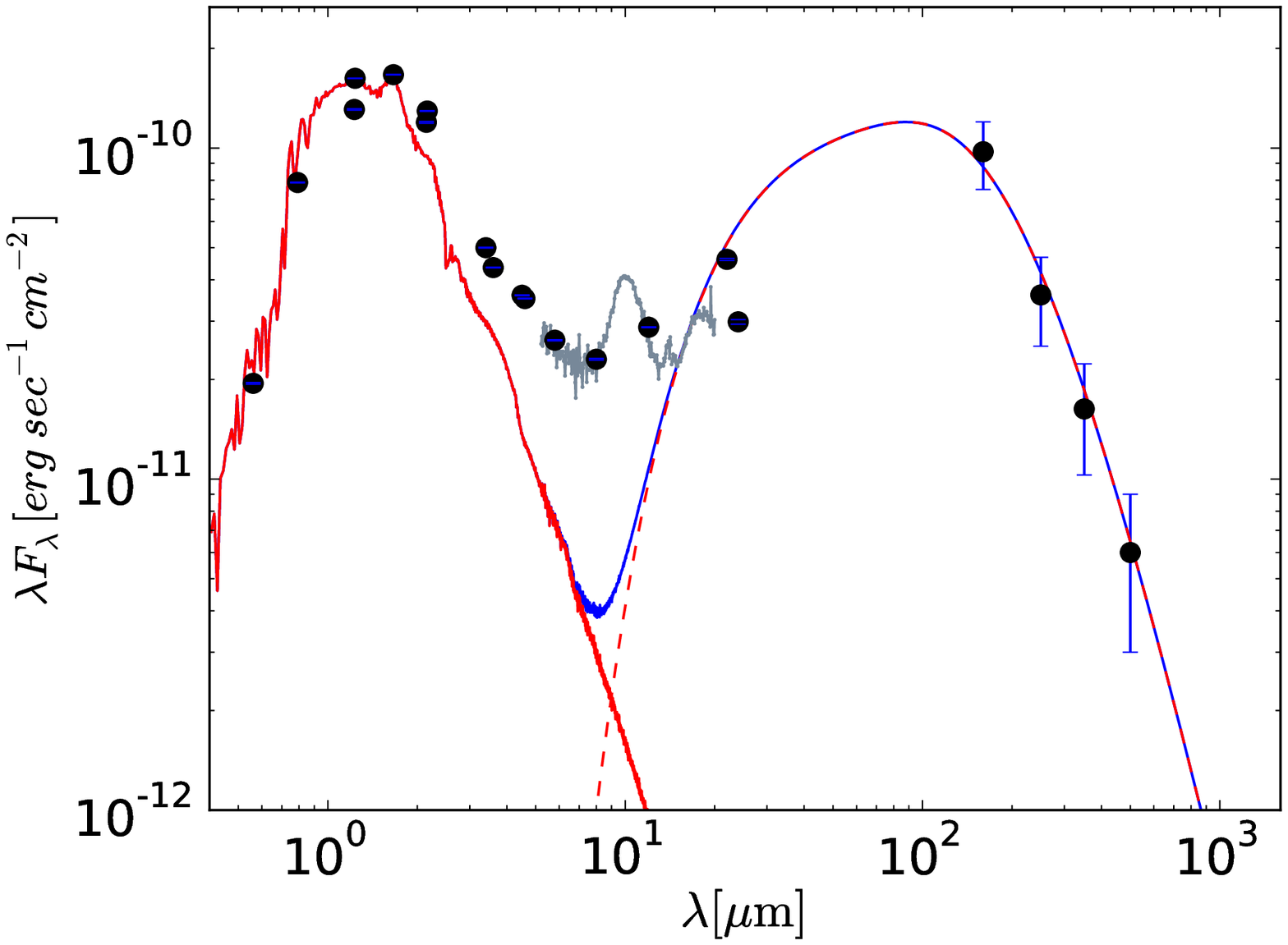}}
\centerline{(a) ISO 237 \hspace{0.25\textwidth} (b) Hn10e }

\caption{SEDs of objects potentially suffering from contamination which are excluded from our sample.\label{fig:contamination}}
\end{figure*}

\citet{manoj_2011}, using Spitzer data, report that ISO 237 had a problem with source contamination. They also report that Hn10e suffered from contamination from a nearby bright object. It is probable that they are also contaminated in the Herschel bands. We exclude ISO 237 and Hn10e from our sample, their SEDs are shown in Fig.~\ref{fig:contamination}. \citet{kim_2009,ribas_2013} do not consider these objects.

\section{Variable sources}
\label{sec:var_sources}

For 7 objects, the photospheric model deviates strongly from some of the optical and NIR datapoints. Since the photometry used to compile the SED spans several decades, the most likely reason for this discrepancy is variability. Whereas long-term red and NIR variations of up to 20\% are common in young stellar objects, objects with more variability are rare \citep{scholz_2012}.

To test the variability hypothesis, we obtained new and simultaneous multi-band images for the 7 objects that we could not fit with the photospheric SED. For this purpose we used the dual-channel imager `Andicam' at the 1.3m telescope on Cerro Tololo. The data was taken in the framework of the SMARTS project DUBLIN-13A-0001 (PI: A. Scholz) over several nights in May 2013.

For all 7 objects we obtained optical images in the R and I bands (3 $\times$ 110 s exposures) and NIR images in the J and K bands (5 $\times$ 30 s in a five-position dither pattern). A standard reduction was carried out including flatfield and bias correction, plus sky subtraction for the NIR data. The magnitudes were measured using aperture photometry. The NIR magnitudes were shifted into the 2MASS system using a few (2-5) other stars in the same field of view. The optical magnitudes were shifted into the Landolt
system using standard star images taken in the same night and at similar airmass. The photometry for these 7 objects is listed in Table ~\ref{table:new_photometry}, in comparison with 2MASS and DENIS magnitudes.

All of them show evidence for variability, which explains the poor fit of the photospheric SEDs. For some of them, the new measurements mostly agree with the DENIS magnitudes, but the 2MASS photometry is inconsistent (see no. 1, 2 or 4 in the table), possibly indicating a short-term event like an eclipse or a burst. One more object (no. 6) has only one major outlier in the DENIS measurements. One star (no. 3) varies within a range of 0.2\,mag in J, but 0.7\,mag in K, indicating that the variability arises in the disc. The remaining two sources (5 and 7) show drastic variations in the J-band (2.3 and 1\,mag), but smaller changes in the K-band (1.2 and 0.4\,mag), i.e. the amplitudes in the J-band are by a factor of $\sim 2$ larger than in the K-band, which can be explained by the presence of accretion-related hot spots or variable extinction along the line of sight \citep{scholz_2009}. For a more detailed assessment of the causes of the
variability, continuous monitoring in multiple bands is required.

\begin{table*}
\centering
\caption{New Andicam photometry for variable sources in comparison with 2MASS and DENIS}

\begin{tabular}{@{}llllllllllll@{}}

\hline
& 2MASS 	      &Name    	          &        &       &        &       & 2MASS          && DENIS 	  &		 \\
& Coordinates	      &	                  &    R   &   I   &  J     &  K    &    J   &  K    &     J 	  &    K	 \\   
\hline
1 & J11075730-7717262 & CHXR30b 	  & 19.9   & 17.6  &  13.10 & 9.59  & 13.87  & 9.95  &  12.9,13.0,13.1 &  9.5,9.8,9.6 \\
2 & J11075792-7738449 & FK Cha, HM16      & 14.38  & 13.30 &  10.40 & 7.25  & 9.50   & 6.83  &  10.4,10.8,10.2 &  7.2,7.5,7.3 \\
3 & J11083905-7716042 & T35, ISO151       & 15.4   & 14.3  &  11.69 & 9.32  & 11.17  & 9.11  &  11.0,10.7      &  9.0,8.9	  \\
4 & J11092266-7634320 & C1-6, P30 	  &  -     &  -    &  -	    & 8.6   & 12.60  & 8.67  &  13.3,12.5,12.6 &  9.4,8.9,8.7 \\
5 & J11092379-7623207 & VZ Cha, T40       & 12.7   & 11.8  &  9.98  & 8.18  & 10.44  & 8.24  &  10.0,9.8,10.1  &  7.8,8.0,8.4 \\
6 & J11095340-7634255 & FM Cha, HM23      & 14.14  & 12.73 &  9.91  & 7.08  & 8.47   & 6.46  &  10.8,10.0      &  6.3,7.5	  \\  
7 & J11113965-7620152 & XX Cha, T49, Sz39 & 13.70  & 12.65 &  10.65 & 9.02  & 10.47  & 8.87  &  10.7,10.7      &  9.5,9.3	  \\

\hline

\end{tabular}
\label{table:new_photometry}
\end{table*}

\subsection{Description of Simple Disc Model}
\label{subsec:model}
The simplest way to model a circumstellar disc is to assume it emits as a series of blackbodies. The model that we used, as in \citet{beckwith_1990}, treats the disc as geometrically thin in the vertical direction. In the radial direction the disc is modelled as a number of concentric rings of dust starting at an inner radius $r_0$ up to an outer radius of $R_d$. This model can reproduce the main features of a SED for a class II source. It is assumed that the dust grains in the disc are predominantly of the same kind and are in thermal equilibrium. The flux density, at frequency $\nu$, is then given by

\begin{equation}
   F(\nu)=\frac{cos(i)}{D^2} \int\limits_{\rmn{r_0}}^{R_d}
   B_\nu(T(r))\left(1 - e^{-\tau_\nu}\right)2\pi r \>
   \rmn{d}r,
\end{equation}
 where $T(r)$ is the temperature of the disc at radius $r$, $B(T_\nu)$ is the Planck function for frequency $\nu$ and $\tau_\nu$ is the optical depth at distance {\it r\/} from the centre of the star for frequency $\nu$. $D$ is the distance to the star from the observer which is taken to be $D = 160$\,pc, $i$ is the angle of inclination of the disc where 90$^\circ$ is edge on. This model fails for high angles of inclination, $ i \gtrsim 80^\circ$, and more sophisticated models are needed, see \citet{chiang_1999}.
 
The temperature and surface density profiles that we adopt are simple power laws, given by:
 
\begin{equation} 
  T(r) = T_0 \bigg(\frac{r}{r_0} \bigg)^{-q}, 
\qquad
 \Sigma(r) = \Sigma_0 \bigg(\frac{r}{r_0} \bigg)^{-p} \\
\label{eq:temp}
\end{equation} 

where $T_0$ and $\Sigma_0$ are the values at $r_0$. The disc mass is given by

\begin{equation}
M_d = \int \limits^{R_d}_{r_0} \Sigma(r)2\pi rdr
\label{eq:M_d}
\end{equation}

We can express the surface density at $r_0$, $\Sigma_0$, in terms of $M_d$ by combining Eqs.~\ref{eq:temp},~\ref{eq:M_d}:

\begin{equation}
\Sigma_0 = \frac{M_d(2-p)}{2\pi r_0^p}(R_d^{2-p} - r_0^{2-p})^{-1}
\end{equation}
The opacity and the optical depth are given by:

\begin{equation}
\kappa_\nu = \kappa_f\bigg( \frac{\nu}{\nu_f}\bigg)^\beta, \nonumber \\
\qquad
\tau_\nu(r) = \frac{\kappa_\nu}{cos \ i} \Sigma(r) 
\quad
\end{equation}

The initial values that we use are $p = 1.5$, $\beta = 1.0$, $q = 0.58$, $r_0 = 5R_{star}$, $R_d = 200$\,AU and $\kappa_{f=\mathrm{230GHz}} = 0.023\,\mathrm{cm}^2 \ \mathrm{g}^{-1}$. We used $T_0 \approx 880(T_*/4000)$\,K following the reasoning of \citet{mohanty_2013} and \citet{andrews_2005}, but with an upper limit at $\sim 1000$\,K. These are the fiducial parameters used in Section~\ref{sec:colour}.

When trying to reproduce observed SEDs (Sections~\ref{sec:disc_masses} and \ref{sec:tds}), some of these parameters were changed within a plausible range. We took $ 0.0001 M_\odot \lesssim M_d \lesssim 0.1 M_\odot$ to be a reasonable range in disc mass, and $50\lesssim R_d \lesssim 200$\,AU for the outer disc radius. We assumed that $ 0.4 \lesssim q \lesssim 0.6$ which is roughly the range given by radiative transfer models \citep{chiang_1997,dalessio_1998} and $0 \lesssim \beta \lesssim 2$ based on observational constraints \citep{ricci_2010}. The parameter $p$ was varied in the range $0.5<p<1.5$. We avoided using $i \gtrsim 80^\circ$, for the reason mentioned above.

\newcommand\aj{AJ} 
\newcommand\actaa{AcA} 
\newcommand\araa{ARA\&A} 
\newcommand\apj{ApJ} 
\newcommand\apjl{ApJ} 
\newcommand\apjs{ApJS} 
\newcommand\aap{A\&A} 
\newcommand\aapr{A\&A~Rev.} 
\newcommand\aaps{A\&AS} 
\newcommand\mnras{MNRAS} 
\newcommand\pasa{PASA} 
\newcommand\pasp{PASP} 
\newcommand\pasj{PASJ} 
\newcommand\solphys{Sol.~Phys.} 
\newcommand\nat{Nature} 
\newcommand\bain{Bulletin of the Astronomical Institutes of the Netherlands}
\newcommand\memsai{Mem. Societa Astronomica Italiana}

\newcommand\apss{Ap\&SS} 
\newcommand\qjras{QJRAS} 

\bibliographystyle{mn2e}
\bibliography{donnabib}

\label{lastpage}

\end{document}